\documentclass[fleqn,usenatbib]{mnras}

\usepackage{newtxtext,newtxmath}
\usepackage[T1]{fontenc}

\DeclareRobustCommand{\VAN}[3]{#2}
\let\VANthebibliography\thebibliography
\def\thebibliography{\DeclareRobustCommand{\VAN}[3]{##3}\VANthebibliography}

\usepackage{graphicx}	% Including figure files
\usepackage{amsmath}	% Advanced maths commands

\usepackage{xcolor}

\title[Disentangling dark and stellar mass]{A Salpeter IMF and an NFW halo: Disentangling the dark and stellar mass of an elliptical galaxy through precise lens modelling of a double-source-plane system}

\author[Tian Li et al.]{
Tian Li\thanks{tian.li@port.ac.uk}$^{1}$,
Thomas E. Collett$^{1}$,
Coleman M. Krawczyk$^{1}$,
Giovanni Granata$^{1}$,
Wolfgang J. R. Enzi$^{1}$,
\and
Daniel J. Ballard$^{2}$,
Natalie E. P. Lines$^{1}$,
Ana Sainz de Murieta $^{1}$,
Luke Weisenbach $^{1}$,
Phil Holloway $^{1}$,
\and
Dan Ryczanowski $^{1}$
\\
$^{1}$ Institute of Cosmology and Gravitation, University of Portsmouth, Burnaby Rd, Portsmouth PO1 3FX, UK\\
$^{2}$ Sydney Institute for Astronomy, School of Physics A28, The University of Sydney, NSW 2006, Australia\\
}

\date{Accepted XXX. Received YYY; in original form ZZZ}
\pubyear{\the\year{}}

\begin{document}
\label{firstpage}
\pagerange{\pageref{firstpage}--\pageref{lastpage}}
\maketitle

\begin{abstract}
We present a strong lensing analysis of the double source plane lens J0946+1006 (colloquially "Jackpot" lens) to measure the inner dark matter density profile, the stellar-to-halo mass ratio, and the stellar initial mass function normalisation using a two component stellar plus dark matter mass model. The stellar mass follows a multi-Gaussian expansion light model with a free global mass-to-light ratio and an allowed radial $M/L$ gradient, while the dark matter is described by an elliptical generalised NFW halo. The double-source-plane geometry provides additional leverage against the mass-sheet transformation and helps constrain the radial mass profile. Despite allowing both a radial stellar $M/L$ gradient and a generalised NFW halo, the data prefer an approximately constant stellar mass-to-light ratio with a Salpeter-like IMF normalisation, and a dark matter halo consistent with NFW. We infer $M_{\star} = 4.4^{+0.25}_{-0.39}\times 10^{11}\,M_{\odot}$ and an inner halo slope $\gamma_{\rm in}^{\rm halo} = 1.04^{+0.10}_{-0.14}$. The halo mass is $M_{200}^{\rm halo} = 1.11^{+0.37}_{-0.32}\times 10^{13}\,M_{\odot}$, implying $\log_{10}(M_{200}/M_{\star})=1.41^{+0.13}_{-0.14}$. At fixed halo mass, the inferred stellar mass lies $\sim0.1$ dex above typical literature stellar halo mass relations at similar redshift, which is comparable to the intrinsic scatter of these relations. We expect this approach to provide a practical template for future dark matter studies with the large double-source-plane lens samples from \textit{Euclid}.

\end{abstract}


\section{Introduction}
Dark matter dominates the mass budget of massive galaxies, yet its distribution is challenging to measure directly because it neither emits nor absorbs light. Its properties must therefore be inferred from its gravitational influence on luminous tracers. Two primary approaches are galaxy kinematics \citep{Rubin1978, Rubin1980} and gravitational lensing \citep[see][for a review]{Treu2010review}. Strong gravitational lensing provides a direct probe of the matter distribution: when two galaxies align along our line of sight, the massive foreground galaxy distorts the light from the background source and forms multiple images or an Einstein ring. Because lensing is sensitive to the total gravitational potential, it constrains the combined distribution of luminous and dark matter.

For a galaxy-scale strong lens, the total projected mass enclosed within the Einstein radius can be measured at the $\sim 1\%$ level \citep{Schneider1992}, and this constraint is largely insensitive to the detailed functional form of the mass distribution. However, this precision is primarily local: strong lensing probes the lens potential at the image positions through the observed lensing configuration and magnification \citep{Birrer2021, O'Riordan2020}. Therefore, translating a precise mass constraint within the Einstein ring into an inference on the global mass profile requires additional assumptions about how mass is distributed with radius.

In the cold dark matter (CDM) paradigm, dark matter haloes are expected to follow universal NFW-like profiles \citep{NFW1997} over a wide range of mass scales \citep{Limousin2007, Caminha2017, Shajib2021}. The stellar mass distribution of massive galaxies is often well described by a S\'ersic profile \citep{Sersic1968}. Remarkably, the superposition of these two components is frequently close to an overall power law in the radial range probed by galaxy-scale strong lenses, a phenomenon known as the bulge-halo conspiracy \citep{True2004, Koopmans2006, Koopmans2009, Auger2010, Dutton2014}.

\begin{figure*}
    \centering
    \includegraphics[width=0.6\textwidth]{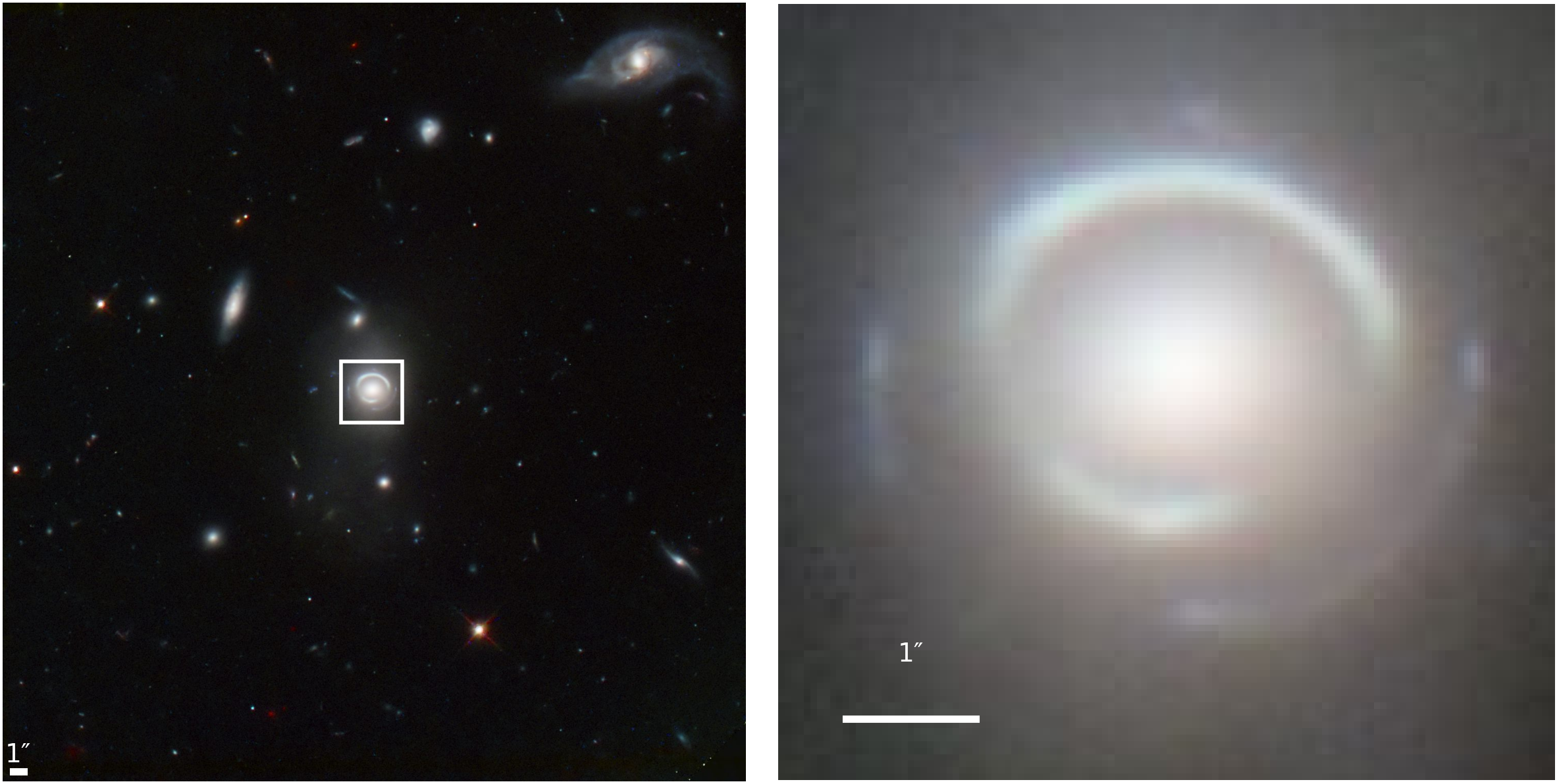}
    \caption{The false-colour image of the Jackpot lens (left), and its $5^{\prime\prime}$ cutout (right). The colour composite is constructed from \textit{HST} bands with \texttt{F475W} mapped to blue (B), \texttt{F606W} to green (G), and \texttt{F814W} to red (R).
}
    \label{fig:rgbimg}
\end{figure*}
This empirical near-power law behaviour motivates a simple and effective modelling choice: the elliptical power law (EPL) mass profile, which has become the industry standard approach for modelling galaxy-scale strong lenses \citep{Shajib2021,Etherington2023, Tdcosmo2025}. The EPL model is also computationally cheap because the deflection angle has analytic expressions \citep{Barkana1998}, which enables fast and robust inference in large scale analyses. To link lensing results more directly to CDM based expectations, and to allow deviations from an exact power law, many studies use composite mass models instead 
\citep{Sonnenfeld2012,Oldham2018, Collett2018, Melo2025}. In these models, the dark matter component is often described by a generalised NFW (gNFW) profile \citep{Keeton2001, Wyithe2001}. In this parametrisation, the inner slope $\gamma_{\rm DM}$ is allowed to vary, so the central profile does not have to follow the standard $r^{-1}$ form.

The inner slope of galaxy dark matter halo encodes both the nature of collisionless CDM and the impact of baryonic processes that can either contract or expand haloes relative to the NFW form. Radiative cooling and in-situ star formation deepen the potential and drive adiabatic contraction, steepening the cusp \citep{Blumenthal1986,Gnedin2004}. In contrast, bursty stellar feedback and AGN-driven outflows can generate rapid potential fluctuations that heat collisionless particles and flatten the central density, producing core-like profiles \citep{PontzenGovernato2012,Martizzi2013}. The balance between these processes is dependent on mass and history: simulations and semi-analytic arguments indicate shallower cusps where feedback dominates at lower masses, with partial recontraction at higher masses as cooling becomes more effective \citep{DiCintio2014a,DiCintio2014b,Tollet2016,Chan2015,Dutton2016}.

Using lens modelling to disentangle the stellar and dark matter components to pin down $\gamma_{\rm DM}$ on kiloparsec scales is challenging because the stellar contribution is set by the mass-to-light ratio ($M/L$), which depends sensitively on the assumed initial mass function (IMF) and is therefore uncertain. Moreover, observations indicate that early-type galaxies (ETGs) often host bottom-heavy IMFs in their central regions \citep{Auger2010}, with radial IMF (and hence $M/L$) gradients that become progressively more Milky Way-like beyond the inner few kpc \citep{vanDokkum2017, Sarzi2018, Sonnenfeld2018, Collett2018, Mehrgan2024}.

Another challenge comes from the main degeneracy inherent in lens modelling, which is the mass-sheet transformation (MST; \citealt[]{Falco1985, Schneider2013}). When a constant mass sheet is added and the convergence is rescaled, the MST can alter the shape of the galaxy mass profile and rescale the size of the source while keeping all lensing observables unchanged. For a composite lens model, this can be achieved through varying stellar $M/L$ gradient and dark matter profile \citep{Li2025a}. Varying the inner slope of the gNFW profile can make the total profile more concave/convex.

Combining lensing and dynamics is usually the way to break this degeneracy. \citet{Shajib2021} analyse a population of SLACS ETGs with joint lensing–dynamics modelling and find that, on average, dark matter haloes are consistent with NFW profiles, with results robust to plausible stellar $M/L$ gradients. \citet{Sheu2025} extend this analysis to a slightly larger sample using hierarchical joint modelling and find a population mean consistent with NFW at low redshift, with a mild trend toward shallower inner slopes at higher redshift. Both studies indicate that DM haloes are close to NFW, but the inferences are limited by modelling assumptions and data quality. Lens light is modelled with single- or double-S\'ersic\ profiles. The lensing constraint is driven by the radial derivative of the deflection angle, and the partial breaking of the mass-sheet degeneracy comes from Jeans modelling; both constraints assumes spherical mass distributions.

In addition to population-level modelling, lens-by-lens analyses have been carried out and have reached differing conclusions about the DM profile. \citet{Sonnenfeld2012,Oldham2018} found very cuspy profiles, whereas \citet{Collett2018,Melo2025} reported DM profiles close to NFW. These studies also have limitations, such as assuming spherical DM haloes, putting ellipticity in the lens potential rather than convergence, and adopting constant stellar $M/L$ ratios. Moreover, all of these studies could be biased by the kinematic assumptions, such as the use of a constant velocity-anisotropy profile, triaxiality and projection effects \citep{Huang2025}.

The MST can also be partially broken in a special and rare class of systems. In these lenses, two background sources at different redshifts are lensed by the same foreground galaxy, often producing Einstein rings at different radii. Such systems are known as compound lenses, or double-source-plane lenses (DSPLs). The MST that is applied on the first source plane will in general fail to reproduce the imaging of the second source, unless an additional mass-sheet term effectively acts with the opposite sign for the second source plane (under fixed cosmological parameters). The detailed derivation of the MST is presented in Appendix \ref{app:mst-dspl}.

The Jackpot lens is the most widely studied DSPL. Figure \ref{fig:rgbimg} shows the false colour image of this lens. \citet{Sonnenfeld2012} jointly modelled \emph{HST} imaging and slit kinematics with a composite mass model and inferred a very steep dark matter inner slope, \(\gamma_{\rm DM}=1.7\pm0.2\). Their lens modelling was relatively simple, and they reduced the DM to a single power-law profile. Using MUSE IFU kinematics, \citet{Turner2024} obtained a consistent result, \(\gamma_{\rm DM}=1.70^{+0.17}_{-0.26}\). Both studies assumed a constant stellar \(M/L\) ratio. 

In this work, we model the Jackpot lens with a star plus dark matter mass profile. The stellar component is represented by a multi-Gaussian expansion (MGE) with a freely varying radial \(M/L\) gradient. The dark matter is described by an elliptical gNFW profile, which we approximate with a 3D MGE of 30 Gaussians to accelerate the computation. Our decomposition is fully auto-differentiable and thus well suited for GPU-accelerated inference.

The paper is organised as follows. In Section~\ref{sec:data}, we describe the data. Section~\ref{sec:modelling} introduces the basics of double-source-plane lensing. In Section~\ref{sec:lensmodel_Setup}, we present our lens-modelling setup. We then report the results under different mass-model assumptions in Section~\ref{sec:lensmodl}, followed by an assessment of systematic uncertainties in Section~\ref{sec:systematics}. Finally, in Section~\ref{sec:discussion}, we discuss the implications of our results and outline future work.
All cosmology-dependent distance ratios (and derived quantities such as $\Sigma_{\rm crit}$) are computed assuming the \textit{Planck} 2018 cosmology \citep{Plank2018}.

\section{Data}
\label{sec:data}
The Jackpot lens is part of the SLACS sample \citep{Gavazzi2008}. Our analysis uses \textsc{HST} observations of J0946+1006 taken with the Advanced Camera for Surveys (ACS) in the F814W, with an exposure time of $t_{\rm exp}=2096\,$s. The image has been drizzled to $0.05^{\prime\prime}$ per pixel \citep[see][]{Collett2014}.
We model the image pixels of the data shown in Figure~\ref{fig:rgbimg}, and
manually create masks for the source arcs to ensure no overlap.

The system consists of a main deflector (hereafter the ``lens'') at
redshift $z_{\rm main}=0.222$ \citep{Gavazzi2008}, and three background sources (S1, S2, and S3).
Their redshifts are $z_{\rm s1}=0.609$ \citep{Gavazzi2008},
$z_{\rm s2}=2.045$ \citep[see][]{Smith2021}, and
$z_{\rm s3} = 5.975$ \citep[see][]{Collett2020}. In this work, our lens modelling uses only the first two source planes (S1 and S2) as the third source is a Lyman-$\alpha$ emitter and is not visible in HST data. We discuss the third source and its impact on the inference in Section~\ref{sec: thirdsource}.

We also obtained data from Multi Unit Spectroscopic Explorer \citep[MUSE,][]{Bacon2010} at the Very Large Telescope (VLT) to measure the line-of-sight stellar velocity dispersion of the Jackpot lens. The adaptive-optics-assisted observations were taken by the ESO Programme ID 0102.A-0950, \citep[PI: Collett, presented in][]{Collett2020} for a total depth of $~5 \, \rm h$.

\section{Modelling the double-source-plane lens}
\label{sec:modelling}

\subsection{Lensing theory}
In this section, we present the theory and conventions/notations used across the paper to describe the lens modelling techniques adopted.
For a lensed image at position $\boldsymbol{\theta}$, the scaled deflection angle of a lens galaxy $\boldsymbol{\alpha}(\boldsymbol{\theta})$ is related to its lensing potential $\psi$ via
\begin{equation}
\boldsymbol{\alpha}(\boldsymbol{\theta})=\nabla \psi(\boldsymbol{\theta})\,,
\end{equation}
and the relation between lensing potential and lensing convergence is
\begin{equation}
\kappa(\boldsymbol{\theta})=\frac{1}{2} \nabla^2 \psi(\boldsymbol{\theta})\,,
\end{equation}
where convergence is defined as
\begin{equation}
\kappa(\boldsymbol{\theta}) \equiv \frac{\Sigma(\boldsymbol{\theta})}{\Sigma_{\mathrm{cr}}}\,.
\end{equation}
The convergence is the lens surface mass density normalised by the critical lensing surface density,
\begin{equation}
\Sigma_{\mathrm{cr}} \equiv \frac{c^2 D_{\mathrm{s}}}{4 \pi G D_{\mathrm{l}} D_{\mathrm{ls}}}\,,
\end{equation}
where \(D\) is the angular diameter distance between two objects, and the subscripts l and s denote the lens galaxy and the source galaxy, respectively.

In a DSPL system, the lens equation of the first source plane can be written as
\begin{equation}
\mathbf{\boldsymbol{\theta}}_1 = \mathbf{\boldsymbol{\theta}}-\, \boldsymbol{\alpha}_1(\mathbf{\boldsymbol{\theta}})\,,
\end{equation}
where \({\boldsymbol{\alpha}_1}\) is the reduced deflection angle of the first source at the image position $\boldsymbol{\theta}$, and \(\boldsymbol{\theta}_1\) is the position of the first source on its source plane. The lens equation of the second source plane is then
\begin{equation}
\boldsymbol{\theta}_2 = \boldsymbol{\theta}-\eta\boldsymbol{\alpha}_1(\boldsymbol{\theta})-\boldsymbol{\alpha}_\mathrm{s1}[\boldsymbol{\theta}-\,\boldsymbol{\alpha}_1(\boldsymbol{\theta})]\,,
\end{equation}
where $\boldsymbol{\theta}_2$ is the position of the second source on its source plane, and $\boldsymbol{\alpha}_\mathrm{s1}$ is the deflection angle of the mass on the first source plane. \(\eta\) is the cosmological scaling factor, which is the ratio of the angular diameter distances between the different redshift planes,
\begin{equation}
\label{eq:beta}
\eta = \frac{D_\mathrm{s1} D_\mathrm{ls2}}{D_\mathrm{ls1} D_\mathrm{s2}}\,,
\end{equation}
where the subscripts s1 and s2 denote source 1 and source 2, respectively. 

In a simple case for which we assume that the lens galaxy has a singular isothermal sphere (SIS) profile, the parameter \(\eta\) is just the ratio of the two Einstein radii. 

As shown by \citet{Schneider2014}, under the assumption that source~2 does not introduce an additional mass sheet (or that any such contribution is negligible), the distance ratio $\eta$ is directly linked to the mass-sheet degeneracy in a DSPL system. In this work, we fix $\eta$ to the value predicted by the Planck18 cosmology, which correspondingly fixes the curvature of the inferred mass profile. The detailed MST relations are given in Appendix~\ref{app:mst-dspl}. As shown in Figure~\ref{fig:mst_eta}, for a DSPL system with $\eta \simeq 1.4$ (as in this work), the MST can be constrained to the $\sim 5\%$ level under typical lens-modelling uncertainties.

\section{Lens modelling strategy}
\label{sec:lensmodel_Setup}
We used the open-source lens modelling code {\tt Herculens}\footnote{\url{https://github.com/Herculens/herculens}} \citep{Galan2022_herculens}, which includes multiplane lensing capabilities (see, e.g. \citealt{Enzi2025}). {\tt Herculens} is built on the automatic differentiation and compilation features of JAX\footnote{\url{https://docs.jax.dev/en/latest/}} and can run on graphics processing units (GPUs). Automatic differentiation efficiently computes partial derivatives without manually deriving or implementing them \citep{Baydin2018}, allowing us to evaluate likelihood gradients with respect to all free parameters.

This section describes our modelling pipeline in four parts: the data, point spread function (PSF) and noise model, the light models (lens and sources), the lens mass model, and the sampling strategy. 
Our fiducial mass model is a composite model consisting of star plus a gNFW dark matter halo (star$+$gNFW). 
To enable a direct comparison with previous composite analyses of the Jackpot (in particular \citealt{Sonnenfeld2012}), we also fit an alternative composite model in which the dark matter halo is represented by an elliptical power-law profile (star$+$EPL).

\subsection{Light model}
\subsubsection{Lens light}
As described in the previous section, we model the lens light as a sum of multiple Gaussian components:
\begin{equation}
\label{eq:MGE_light}
F_{\mathrm{light}}(x,y) \;\approx\; \sum_{n=1}^{N}
A_{\mathrm{light},n}\,
\exp\!\left[-\frac{R^2}{2\,\sigma_n^{2}}\right],
\end{equation}
where, for each Gaussian component, the amplitude $A_{\mathrm{light},n}$, the axis ratio $q_n$ ($R^2 \equiv q_n\,(x-x_n)^2 + \frac{(y-y_n)^2}{q_n}$), the centroid $(x_n,y_n)$, and the ellipticity ($e_1, e_2$) are treated as independent free parameters. We adopt $N=15$ Gaussians with dispersions $\{\sigma_n\}$ uniformly spaced in $\log \sigma_n$ between $0.001$ and $3$~arcsec. The Einstein radius measurement is self-consistent when using 15 or more components.

We parametrise the ellipticity using the Cartesian components $(e_1,e_2)$ rather than the axis ratio $q$ and position angle $\phi$. This avoids the $\pi$-periodicity of $\phi$ and provides a continuous parameter space, which improves the efficiency and robustness of the inference. We adopt the standard mapping:
\begin{equation}
\label{eq:e1e2_def}
e_{1,n} = \frac{1-q_n}{1+q_n}\cos(2\phi_n), 
\qquad
e_{2,n} = \frac{1-q_n}{1+q_n}\sin(2\phi_n).
\end{equation}

\subsubsection{Source light parametric}

We begin by modelling the source surface brightness with an MGE, analogous to our treatment of the lens light. For the Jackpot lens, this parametrisation provides an excellent fit to the F814W data: the light of source~1 is dominated by smooth continuum emission with minimal clumpiness. We model both source~1 and source~2 using four elliptical Gaussian components. The Gaussian scale radii are assigned to four logarithmically spaced bins between $0.001$ and $0.2$ arcsec, and each component draws its scale radius from a log-uniform prior within its assigned bin.

\subsubsection{Source light pixelated}
We then model the sources as fields defined on regular Cartesian pixel grids that can be treated as a Gaussian process (GP). For each source, the grid is chosen as the smallest square that encloses all mask pixels traced to the source plane. To mitigate overextension of the mask (e.g., due to the PSF or mapping-induced outliers in the source plane), we shrink the side length of the square by factors of 0.7 and 0.6 for source~1 and source~2, respectively.

The field value in each pixel forms an element of the vector $\vec{s}$. In Fourier space, the mode amplitudes can be determined by a power spectrum \citep[see, e.g., ][]{Galan2024, RustigEtAl2024_LensCharm}:
\[
\vec{s} \;=\; \mathcal{F}^{-1}\!\big[\,\sqrt{P}\,\odot\,\vec{\xi}\,\big],
\]
where \(P\) is the Matérn power spectrum evaluated on the Fourier grid, 
\(\odot\) denotes element-wise multiplication, and $\vec{\xi}$ is standard Gaussian white noise. The Matérn power spectrum \citep[see e.g.][]{Stein2012Interpolation} corresponding to a Matérn covariance kernel is
\begin{equation}
P(k)=\sigma^{2}\,4\pi n\left(\frac{2n}{\rho^{2}}\right)^{n}\left(\frac{2n}{\rho^{2}}+k^{2}\right)^{-(n+1)},
\end{equation}
where $\sigma$ sets the overall amplitude, $\rho$ is the correlation length, and $n$ controls the smoothness. We also fit for these parameters using Jeffery's prior during the inference. 

Applying the discrete inverse Fourier transform \(\mathcal{F}^{-1}\) yields the pixelated source image \(\vec{s}\), which is a realisation of a Gaussian process with Matérn covariance.

To enforce non-negative source-plane surface brightness while keeping the model differentiable, we apply a smooth positivity map to the latent pixel field,
\begin{equation}
\vec{s}\;\longrightarrow\;\frac{\mathrm{softplus}(h\,\vec{s})}{h},\qquad 
\mathrm{softplus}(x)=\log\!\left(1+e^{x}\right),\qquad h = 100.
\end{equation}
The softplus function is strictly positive and smooth: for \(x\ll 0\), \(\mathrm{softplus}(x)\approx e^{x}\) so it approaches \(0\) exponentially, while for \(x\gg 0\), \(\mathrm{softplus}(x)\approx x\) and thus becomes approximately linear. The scale parameter \(h\) controls how sharply the mapping transitions from near-zero values for negative inputs to an approximately identity mapping for positive inputs.

We use $50\times50$-pixel grids for both sources, which provide sufficient ray-tracing samples per pixel and satisfy the Nyquist criterion: on average, four image-plane pixels map onto one source pixel.
The priors used for lens and source light is presented in Table \ref{tab:prior_all}.

\subsubsection{Lens image and noise model}

Our analysis uses a PSF derived from a nearby field star (following \citealt{Collett2018}). The noiseless image model is the sum of all light components (the lens galaxy and the lensed sources) convolved with the PSF. We compute this convolution in Fourier space using FFT-based multiplication, and apply zero-padding to suppress wrap-around artefacts.

For the image-noise map $\sigma_{\rm img}$, we estimate the background level from the root-mean-square of blank-sky pixels, $\sigma_{\rm bg}$, and include a free rescaling parameter with a uniform prior in the range $0.8$--$1.2$ to account for possible inaccuracies in this estimate. We compute the Poisson (shot) noise term, $\sigma_{\rm shot}$, from the model image under Poisson statistics. The total noise map is then taken as the quadrature sum
\[
\sigma_{\rm img} \;=\; \sqrt{\sigma_{\rm bg}^{2} + \sigma_{\rm shot}^{2}}\,.
\]

We model the data at the native drizzled pixel scale, which is sufficient to fit the arcs to the noise level and to constrain the deflection field at the image positions. We discuss potential systematics associated with finite sampling and unresolved substructure in Section~\ref{sec:systematics}.

\subsection{Mass Model}

Our mass model includes an MGE stellar mass component and a gNFW profile for the DM halo mass component approximated with a 3D MGE of 30 Gaussians. 

\begin{table*}
\centering
\small
\renewcommand{\arraystretch}{1.2}
\setlength{\tabcolsep}{6pt}
\begin{tabular}{lcc}
\hline
\textbf{Parameter / prior} &
\textbf{STAR + EPL} &
\textbf{STAR + gNFW} \\
\hline

\multicolumn{3}{l}{\emph{Main lens mass (stars + dark matter)}}\\
Einstein radius / convergence
& $\theta_{\rm E}\sim \mathcal{U}(0.01,\,2.0)$
& $\kappa_s \sim \mathcal{U}(0,\,1)$ \\
Density slope
& $\gamma \sim \mathcal{U}(0.1,\,3.0)$
& $\gamma_{\rm in} \sim \mathcal{U}(0.01,\,3.0)$ \\
Ellipticity $(e_1,e_2)$
& $\mathrm{TN}(0,\,0.25;\,-0.4,\,0.4)$
& $\mathrm{TN}(0,\,0.25;\,-0.4,\,0.4)$ \\
External shear $(\gamma_1,\gamma_2)$
& $\mathcal{U}(-0.2,\,0.2)$
& $\mathcal{U}(-0.2,\,0.2)$ \\
Stellar mass scale
& $\mathcal{U}(0,\,14)$
& $\mathcal{U}(0,\,14)$ \\
Stellar M/L gradient
& $\mathcal{U}(-0.6,\,0)$
& $\mathcal{U}(-0.6,\,0)$ \\
Scale radius
& --
& $r_s \sim \mathcal{U}(2.5,\,20)$ \\
\hline

\multicolumn{3}{l}{\emph{Source-plane mass (Source 1)}}\\
Einstein radius (SIS)
& $\theta_{\rm E}\sim \mathcal{U}(0.05,\,0.3)$
& $\theta_{\rm E}\sim \mathcal{U}(0.05,\,0.3)$ \\
\hline

% ---------------------------
% Shared light priors
% ---------------------------

\multicolumn{3}{l}{\emph{Lens light (Gaussian components)}}\\
Amplitude scale &
\multicolumn{2}{c}{$\mathrm{LogU}(10^{-5},\,10^{4})$} \\
Scale radius &
\multicolumn{2}{c}{$\mathrm{LogU}(\sigma^{\rm bin}_{j-1},\,\sigma^{\rm bin}_{j}),\ 
\sigma^{\rm bin}\in[0.001,3]$} \\
Centroid &
\multicolumn{2}{c}{$\mathrm{TN}(0,\,0.1;\,-0.2,\,0.2)$ } \\
Ellipticity &
\multicolumn{2}{c}{$\mathrm{TN}(0,\,0.1;\,-0.2,\,0.2)$ } \\

\hline

\multicolumn{3}{l}{\emph{Source light (parametric)}}\\
Amplitude scale &
\multicolumn{2}{c}{$\mathrm{LogU}(10^{-5},\,10^{4})$} \\
Scale radius &
\multicolumn{2}{c}{$ \mathrm{LogU}(\sigma^{\rm bin}_{j-1},\,\sigma^{\rm bin}_{j}),\ 
\sigma^{\rm bin}\in[0.001,\,0.2]$} \\
Centroid  &
\multicolumn{2}{c}{$\mathcal{N}(0,\,0.5^2)$ } \\
Ellipticity &
\multicolumn{2}{c}{$\mathcal{N}(0,\,0.1^2)$ } \\

\multicolumn{3}{l}{\emph{Source light (pixelated)}}\\
Spectral index / smoothness $n$ &
\multicolumn{2}{c}{$n \sim \mathrm{TW}(-1,\,10^{-4};\,n_{\rm high}=100)$} \\
Amplitude $\sigma$ &
\multicolumn{2}{c}{$\mathrm{LogU}(10^{-5},\,10)$} \\
Correlation scale $\rho$ &
\multicolumn{2}{c}{$\rho \sim \mathrm{LogNormal}(2.1,\,1.1)$} \\
Fourier white noise modes &
\multicolumn{2}{c}{$\epsilon_{\boldsymbol{k}} \sim \mathcal{N}(0,\,1)$ } \\
\hline

\end{tabular}
\caption{
Priors for all mass, light, and pixelated-source parameters used in this work.
Light priors are identical for the two composite mass models (STAR+EPL and STAR+gNFW).
$\mathcal{U}$ denotes a Uniform prior, $\mathrm{LogU}$ a log-uniform prior, $\mathcal{N}$ a Normal prior, $\mathrm{TN}$ a truncated Normal, and $\mathrm{TW}$ the truncated-wedge prior used for the Mat\'ern index.
The Gaussian scale radii are drawn from log-uniform priors within logarithmically spaced bins:
the bin edges $\{\sigma^{\rm bin}_k\}_{k=0}^{N_{\rm G}}$ are log-spaced between the quoted limits, and each component draws $\sigma_j$ from its assigned interval $(\sigma^{\rm bin}_{j-1},\sigma^{\rm bin}_{j})$.
}
\label{tab:prior_all}
\end{table*}

\subsubsection{Stellar mass component}
Our fiducial model consists of stars plus dark matter. The stellar mass distribution is tied to the multi-Gaussian light model: the amplitude of each Gaussian is linked to its contribution to the stellar convergence, such that the relative convergence amplitudes between Gaussians are fixed by the light model. An $M/L$ ratio is required to map the light model to a mass model; here we treat the (global) $M/L$ as a free parameter. Following \citet{Oldham2018}, we also allow for a radial $M/L$ gradient:
\begin{equation}
\Upsilon_{\star}(\sigma_i) \propto \sigma_i^{-\nabla M/L},
\end{equation}
where $\Upsilon_{\star}$ is the mass-to-light ratio assigned to each Gaussian component, $\sigma_i$ is the standard deviation of each Gaussian, and $\nabla M/L$ is the slope of the $M/L$ gradient. Here $\nabla M/L$ is an additional free parameter that controls a smooth, global radial variation of the stellar $M/L$ across the Gaussian components; $\nabla M/L=0$ corresponds to a constant $M/L$.

We normalise the stellar-mass MGE such that the Gaussian amplitudes sum to unity in convergence. For the Jackpot lens this normalisation implies a mass scale of $4.8\times10^{10}\,M_\odot$ per arcsec$^{2}$, so the stellar mass enclosed within the Einstein radius is $4.8\times10^{10}\,M_\odot \times \pi \theta_{\rm E}^{2}$. In practice, this means we do not explicitly parametrise the stellar component with an $M/L$ in physical units ($M_\odot\,{\rm pc}^{-2}$). Instead, we scale the normalised Gaussian components by a free parameter $\Upsilon_\kappa$ and apply the $M/L$ gradient. The amplitude of each Gaussian is then expressed as:
\begin{equation}
\Upsilon_{\kappa,i}(\sigma_i)=
\Upsilon_{\kappa}\,
\frac{A_{{\rm light},i}}{\sum_{j=1}^{N} A_{{\rm light},j}}\,
\sigma_i^{-\nabla M/L},
\end{equation}
where $\Upsilon_{\kappa,i}$ is the mass amplitude of the $i$th Gaussian.

\subsubsection{Dark matter mass component}
The 3D radial profile of the dark matter is described as a elliptical gNFW profile parametrised as:
\begin{equation}
\rho(R)=\rho_s\left(\frac{R}{r_s}\right)^{-\gamma_{\rm in}}
\left(1+\frac{R}{r_s}\right)^{\gamma_{\rm in}-3},
\end{equation}
where $\rho_s$ is the characteristic density at the scale radius $r_s$, and
$\gamma_{\rm in}$ is the inner density slope (cuspy if $\gamma_{\rm in}>1$; cored if $\gamma_{\rm in}=0$).
For $\gamma_{\rm in}=1$, the profile reduces to the classical NFW.

The gNFW profile is approximated using a 3D MGE with 30 components, i.e. the dark matter mass distribution is represented as a sum of concentric elliptical Gaussian basis functions \citep{Emsellem1994_MGE, Cappellari2002_MGE, Shajib2019MGE}. A full description of the 3D MGE implementation is provided in Appendix~\ref{app:jaxlensingprofile}.

\subsubsection{Elliptical Power-Law}
Traditional strong-lens analyses often describe the total projected mass distribution with an EPL model \citep[e.g.][]{Collett2014}. The convergence ($\kappa_{\mathrm{EPL}}$) of the EPL model can be parametrised as
\begin{equation}
\kappa_{\mathrm{EPL}}(R, \gamma, q)=\frac{3-\gamma}{2}\left(\frac{\theta_\mathrm{E}}{R}\right)^{\gamma-1}\,,
\end{equation}
where $\gamma$ is the logarithmic density slope and $\theta_\mathrm{E}$ is the Einstein radius. For $\gamma=2$, the model reduces to the singular isothermal case (SIS/SIE). In our DSPL modelling, the mass associated with the second deflector (the first source plane) is modelled as an SIS.

In this work, the EPL model is not used as our  mass parametrisation. Where relevant, we take the EPL parameter reported by \citet{Collett2014} as a reference.

For a direct comparison with \citet{Sonnenfeld2012}, we replace the gNFW halo in our fiducial composite model with an EPL profile as an approximation to the dark matter halo (star$+$EPL).

\begin{figure*}
    \centering
    \includegraphics[width=0.9\textwidth]{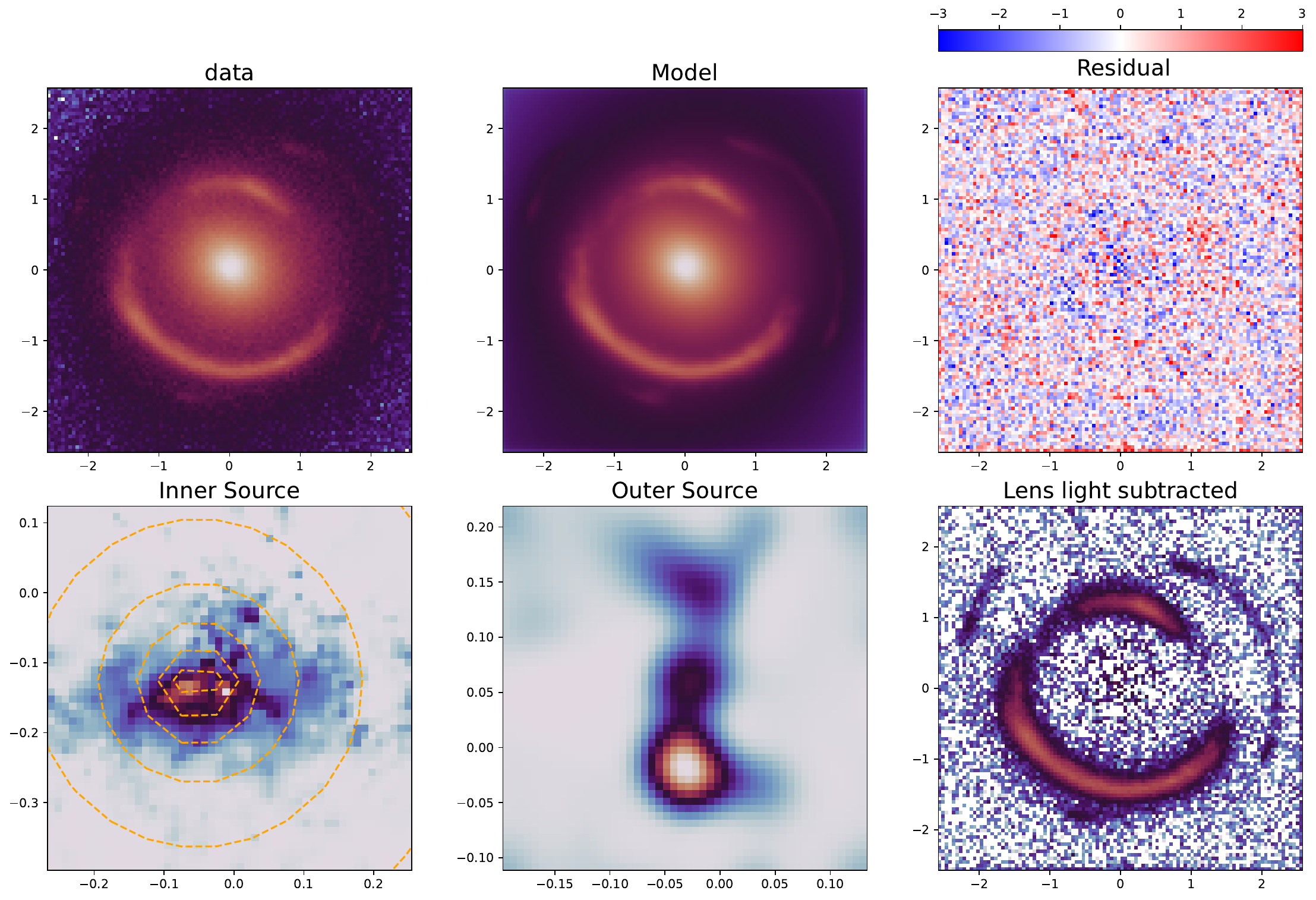}
    \caption{The lens model of star + gNFW mass model. \textit{Top-left:} observed image. \textit{Top-middle:} best-fit model image (posterior-median). \textit{Top-right:} normalized residual $(\mathrm{data}-\mathrm{model})/\sigma$, clipped to $\pm 3\sigma$. \textit{Bottom-left:} reconstructed inner source $s_1$ on its source plane; orange dashed contours show $\log\kappa$ of the $s_1$ mass component. \textit{Bottom-middle:} reconstructed outer source $s_2$ on its source plane. \textit{Bottom-right:} lens-light subtracted image with the same log stretch as the top-left panel. Axes are in arcsecond.}

    \label{fig:lensmodel}
\end{figure*}

\subsection{Initialisation}

Lens modelling of this data consists of $\sim 3000$ free parameters (\(90\) for the lens light, 12 for the mass model, 3 for the source power spectrum and 2500 for the source pixels), to make the Markov chain Monte Carlo (MCMC ) chain converge to the correct posterior, we must provide a starting point that is close enough to the true value.

We used the {\tt NumPyro} implementation of stochastic variational inference \citep[SVI; see][]{WingateWeber2013}. {\tt NumPyro} is built on JAX\footnote{\url{https://github.com/google/jax}} for efficient automatic differentiation. SVI is a algorithm that approximates the posterior with a tractable variational distribution by optimising its parameters using stochastic gradients. We employed the AdaBelief optimiser \citep{Zhuang2020} with a low-rank multivariate normal as the guiding (variational) distribution. This guide is not flexible enough to capture the full posterior in our $\sim 3000$-dimensional problem, but it is much cheaper than MCMC because SVI optimises variational parameters by minimising the Kullback--Leibler divergence to the true posterior, rather than drawing posterior samples.

Since the inner arcs of our DSPL candidates are blended with the lens light, we model the lens light and the arcs simultaneously. We first describe each source with four elliptical Gaussian luminosity profiles to obtain an initial approximation. After this parametric SVI stage, we fit the source on a pixel grid to obtain the Matérn power-spectrum and corresponding noise field of this image, and then use that fit as the initialisation for the pixelated SVI. After the pixelated SVI, the lens model is already close to the final result, except that the source is still relatively smooth and the mass model posterior is still too simple.

% For the parametric model 8 independent SVI chains are initialized from the prior distribution and run from 10000 iterations. These chains are used as the starting point for the pixelated SVI optimization which is run for 50000 iterations. This lens light of the chain with the lowest average loss value (corresponding the the largest ELBO) is chosen as our fiducial lens light. The lens light is then fixed during HMC to reduce the computation cost.

For the parametric lens-light model, we run eight independent SVI chains, each initialised from the prior, for 10{,}000 iterations. These solutions are then used to initialise the pixelated SVI optimisation, which is run for 50{,}000 iterations.

SVI maximises the evidence lower bound (ELBO). We adopt a low-rank multivariate normal guide. Optimisation is performed with the AdaBelief optimiser and a two-stage exponential learning-rate schedule. Specifically, the learning rate is initialised to $10^{-2}$ and decays exponentially with decay rate 0.99 and transition steps 200 for the first half of the optimisation. At iteration 25{,}000 we switch to a second exponential decay with the same decay rate (0.99) but transition steps 10, initialised to the learning rate at the boundary to ensure continuity.

We select the chain with the lowest mean SVI loss (i.e. the highest ELBO) and adopt its lens-light parameters as our fiducial lens-light model. The lens light is then held fixed during HMC sampling to reduce computational cost.

\begin{figure*}
    \centering
    \includegraphics[width=0.8\textwidth]{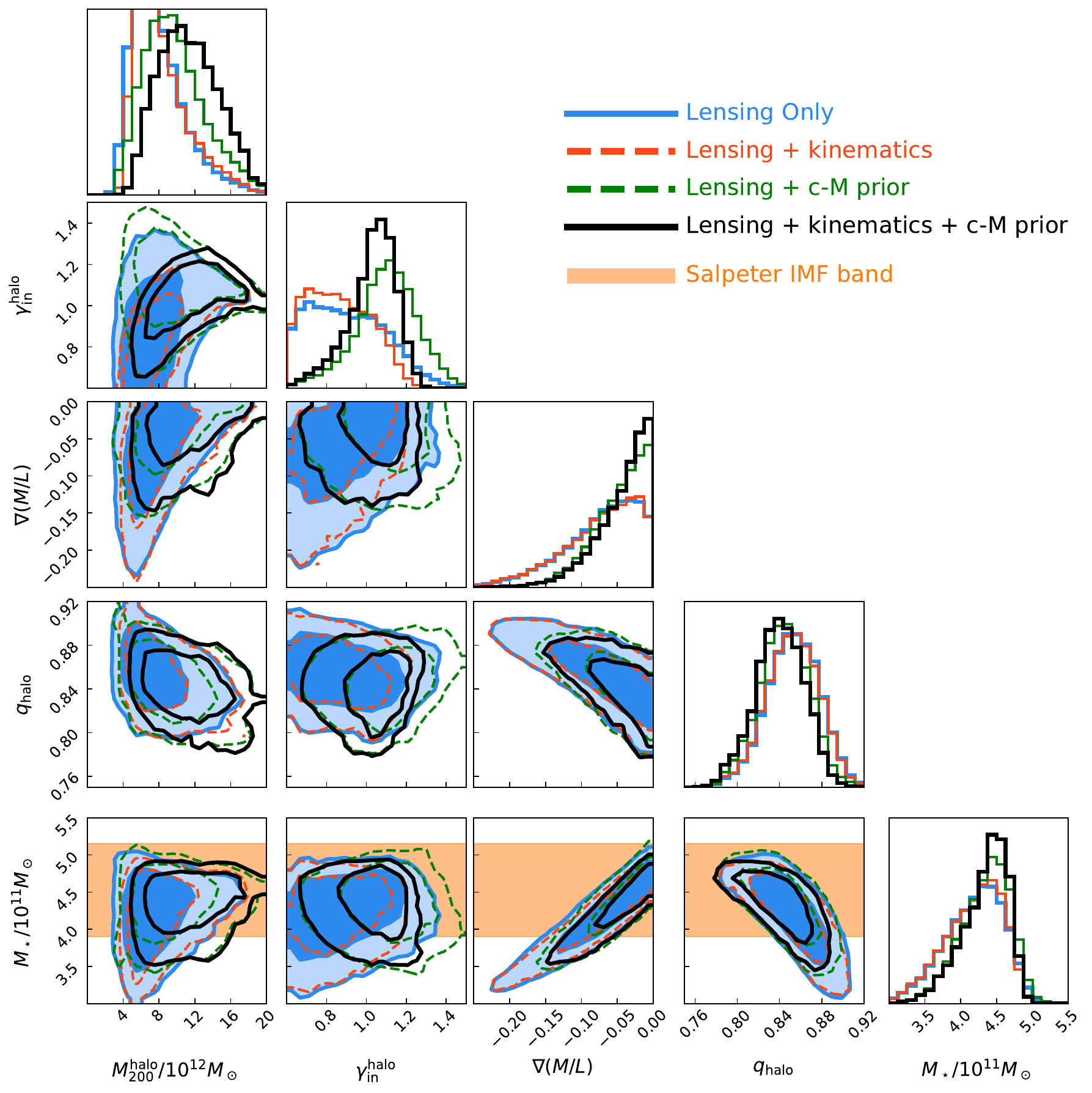}
    \caption{Posterior distributions for the composite star + gNFW halo model under different combinations of external information.
    Diagonal panels show the 1D marginalised posteriors; off-diagonal panels show the 2D posteriors with 68\% and 95\% credible contours.
    The parameters shown are $M_{\mathrm{200}}^{\mathrm{halo}} / 10^{12} M_\odot$ (halo mass),
    $\gamma_{\mathrm{in}}^{\mathrm{halo}}$ (inner logarithmic density slope),
    $\nabla(M/L)$ (logarithmic radial gradient of the stellar mass-to-light ratio),
    $q_{\mathrm{halo}}$ (projected axis ratio),
    and $M_\star / 10^{11} M_\odot$ (total stellar mass).
    Filled blue contours show the lensing-only inference; orange dashed lines show the posterior reweighted by the kinematic likelihood;
    green dashed lines include the concentration-mass prior; and solid black lines show the combined lensing + kinematic + concentration-mass constraints.
    The shaded horizontal band marks the stellar-mass range corresponding to a Salpeter IMF which comes from \citet{Sonnenfeld2012}.}
    \label{fig:posterior}
\end{figure*}

\subsection{BAYESIAN INFERENCE}

We estimate the posterior with NumPyro’s implementation of the No-U-Turn Sampler (NUTS; \citealp{HoffmanGelman2014NUTS,Phan2019NumPyro}), a Hamiltonian Monte Carlo (HMC) method \citep{1987PhLB..195..216D,brooks2011handbook} that uses gradients to propose efficient moves. Compared with traditional MCMC, HMC/NUTS typically yields low–autocorrelation chains, needs fewer warm-up steps, and scales well to high-dimensional posteriors. To further improve robustness, we combine NUTS with an HMC-within-Gibbs scheme \citep{Krawczyk2024MultiHMCGibbs}: we update blocks of parameters in turn while holding the others fixed, following a Gibbs-like schedule \citep[cf.][]{GelmanBDA3}. We split the sampler into two Gibbs steps: one updates all light parameters, and the other updates all mass parameters.

For both the SVI initialization and the HMC-within-Gibbs stage, we run eight independent chains with distinct random seeds. For our fiducial model, we draw 20{}000 posterior samples per chain after 10{}000 warm-up steps. For the simpler EPL model, we use 4{}000 warm-up steps and draw 2{}000 posterior samples per chain. The typical $\hat{r}$ for each fit parameter is smaller than 1.05 (mostly around 1.02), indicating that all the chains independently converged on the same posteriors. While there are some divergent transitions during sampling, they occur in fewer than 0.1$\%$ of the draws, which is low enough that they should not bias the results.

Table \ref{tab:prior_all} lists the priors used in our HMC analysis. The slope of the stellar mass-to-light gradient is assigned a uniform prior on $[-0.6,0]$, where $\nabla M/L=-0.6$ approximates an IMF that transitions from Salpeter-like in the centre to Chabrier-like in the outskirts. To set a fiducial scale for the NFW halo scale radius, we adopt the empirical size--halo relation
$R_{\mathrm{e}} = k R_{200\mathrm{c}}$
appropriate for ETGs, with $k=0.012$ \citep[e.g.][]{Kravtsov2013,Huang2017}, together with a representative concentration
$c_{200}=8.11$
from a standard $c(M,z)$ relation \citep[e.g.][]{DuttonMaccio2014,DiemerJoyce2019}. For the Jackpot lens, this gives

\[
R_{\mathrm{s},0}
= \frac{R_{200\mathrm{c}}}{c_{200}}
= \frac{R_{\mathrm{e}}}{k\,c_{200}}
\approx \frac{0.5^{\prime\prime}}{0.012\times 8.11}
\approx 5.14^{\prime\prime},
\]

which is close to the commonly adopted heuristic $R_{\mathrm{s}}\sim 10\,R_{\mathrm{e}}$ \citep[e.g.][]{Sonnenfeld2023,Melo2025}. To estimate a plausible interval around this fiducial value, we propagate independent log-normal scatter in $k$ and $c_{200}$ through

\[
\log_{10} R_{\mathrm{s}}
=
\log_{10} R_{\mathrm{e}}
-
\log_{10} k
-
\log_{10} c_{200},
\]

so that

\[
\sigma_{\log_{10} R_{\mathrm{s}}}
=
\sqrt{\sigma_{\log_{10} k}^{2}+\sigma_{\log_{10} c}^{2}},
\]

where $\sigma_{\log_{10} k}$ and $\sigma_{\log_{10} c}$ denote the scatter in $\log_{10} k$ and $\log_{10} c_{200}$, respectively. We adopt $\sigma_{\log_{10} k}=0.25$ dex as a conservative allowance for the spread in ETG size--halo calibrations, motivated by the $\sim 0.2$--$0.3$ dex variation discussed by \citet{Huang2017}, and $\sigma_{\log_{10} c}=0.18$ dex for the intrinsic scatter in the concentration--mass relation. This gives
$
\sigma_{\log_{10} R_{\mathrm{s}}}
\approx 0.31~{\rm dex}.
$
The corresponding $1\sigma$ interval is therefore
$
R_{\mathrm{s}}
\approx
2.5^{\prime\prime}\text{--}10^{\prime\prime},
$
while the approximate $2\sigma$ interval is
$
R_{\mathrm{s}}
\approx
1^{\prime\prime}\text{--}20^{\prime\prime}.
$

Because the lower end of the $2\sigma$ interval corresponds to an extremely compact halo relative to the ETG-motivated fiducial scale, we adopt the lower bound from the $1\sigma$ interval, $2.5^{\prime\prime}$, and the upper bound from the broader $2\sigma$ interval, $20^{\prime\prime}$. We therefore use the sampling prior
$R_{\mathrm{s}} \sim \mathcal{U}(2.5,20)\,\mathrm{arcsec}$.

The lens model results including the modelled image, residual and two sources are presented in Figure \ref{fig:lensmodel}. The posterior distribution of the lensing only result is shown as the blue contour in Figure \ref{fig:posterior}, and the full posterior is shown in Figure \ref{fig:full_posterior}

\subsection{Importance Sampling with Kinematic and Concentration Relation}
\begin{figure*}
    \centering
    \includegraphics[width=0.6\textwidth]{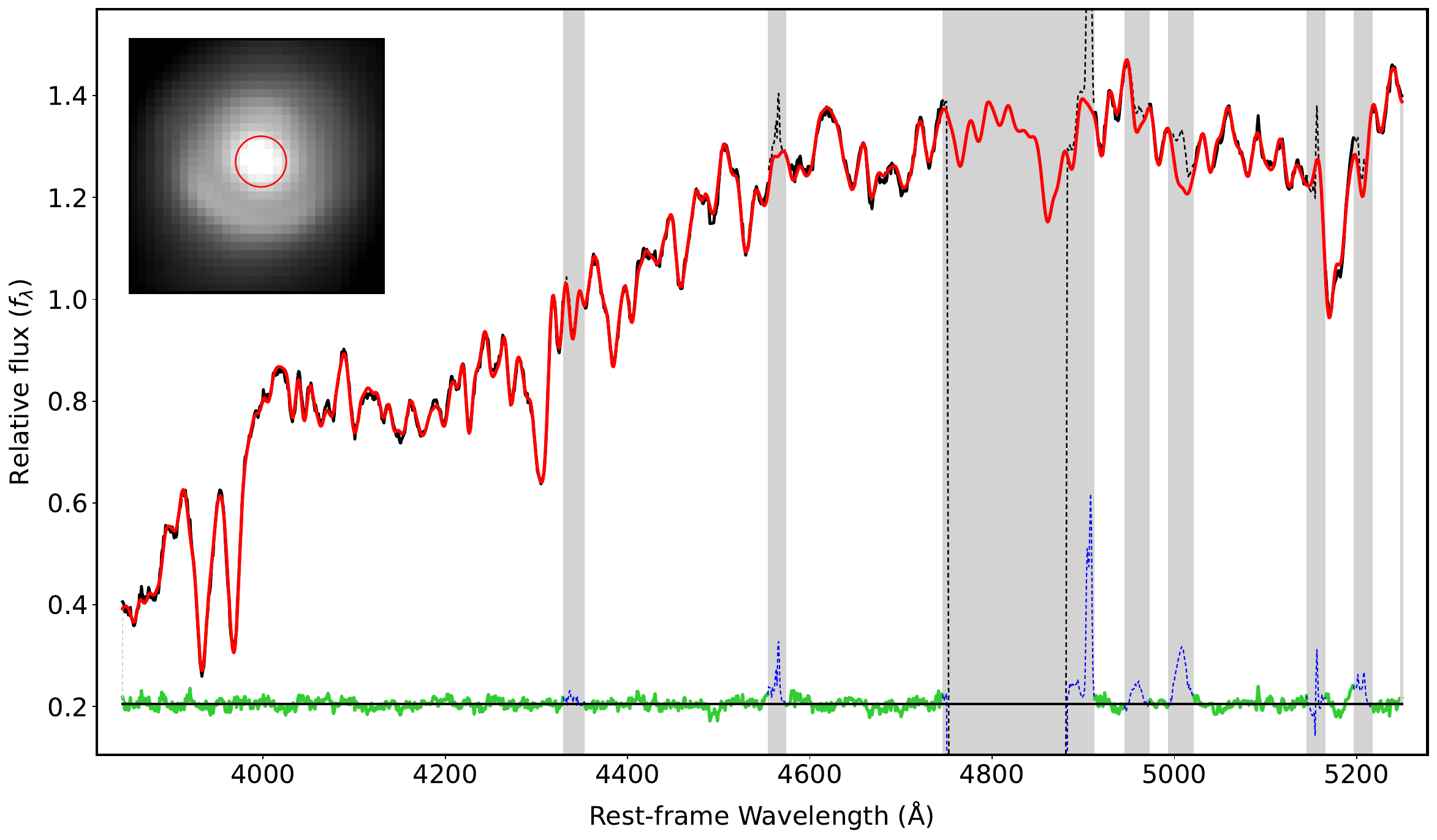}
    \caption{Aperture-averaged line-of-sight stellar velocity dispersion fit with \texttt{pPXF}. The observed MUSE spectrum and the \texttt{pPXF} best-fit model (normalised to a median flux of 1) are marked in black and red, respectively. The residuals of the data with respect to the model are shown in green. The regions masked, to avoid the impact of the lensed source emission lines or of the subtraction of sky, laser, or telluric lines, are shaded in grey. The data minus model residuals in these regions are marked in blue. In the top-left corner, we show a cutout of the summed white image of the MUSE cube, marking the circular aperture used to extract the lens spectrum in red.}
    \label{fig:kin}
\end{figure*}

To assess whether our posterior distribution can be further constrained by kinematic information, we take advantage of integral-field spectroscopic data from the MUSE to measure the line-of-sight stellar velocity dispersion of the Jackpot lens. The $~5 \, \rm h$ integration allows us to extract a high signal-to-noise spectrum $\left(S/N = 75.6\,\text{\AA}^{-1}\right)$ within a circular aperture of radius $0.6^{\prime\prime}$. We measure the line-of-sight stellar velocity dispersion $\sigma_{\rm los}^{\rm obs}$ using the public spectral fitting code \texttt{pPXF}\footnote{Version 8.2.6} \citep[penalized pixel-fitting;][]{Cappellari2004,Cappellari2023} to perform a full-spectrum fit of the continuum and of the absorption lines of the observed spectra in the rest-frame wavelength range $[3850-5250]\, \text{\AA}$. We fit the stellar population and line-of-sight velocity distribution simultaneously by matching the galaxy spectrum with a curated subset of 462 high-resolution X-shooter Spectral Library (XSL DR3) templates, selected following \citet{Knabel2025}, finding $\sigma_{\rm los}^{\rm obs}=249.1 \pm 2.1~\mathrm{km\,s^{-1}}$. We show the results of our \texttt{pPXF} spectral fit in Figure \ref{fig:kin}, which shows the very small data-to-model residuals achieved.

% The details of the \texttt{pPXF} set-up (e.g. the evaluation of the uncertainty on $\sigma_{\rm los}$ and the masking of spurious features from the subtraction of of sky, laser, or telluric lines) are reported in Granata et al. (submitted).

As we wish to evaluate how well our lens mass models agree with the observed value of the stellar velocity dispersion, we take advantage of the inferred total mass and light distributions to evaluate a theoretical line-of-sight velocity dispersion. Because each of Gaussian components in the lens-light measurements are non-concentric and have different position angles, we use the axis-symmetric Jeans modelling code \textsc{JamPy} \citep{Cappellari2008} to compute the predicted line-of-sight velocity dispersion in a spherical aperture for each mass model in the posterior of our fiducial lens model. The resulting theoretical velocity dispersions are shown in the bottom panel of Figure~\ref{fig:full_posterior}. For each posterior sample we assign a kinematic weight
\begin{equation}
    w_{\rm kin} \propto \exp\left[
        -\frac{1}{2}
        \left(
            \frac{\sigma_{\rm los}^{\rm model} - \sigma_{\rm los}^{\rm obs}}
                 {\delta\sigma_{\rm los}^{\rm obs}}
        \right)^{2}
    \right],
\end{equation}
and construct a new, reweighted posterior labelled “lensing + kinematic”. This reweighted posterior is shown in orange dashed line in Figure \ref{fig:posterior}.

In addition, we incorporate information from the $\mathrm{c}$–$M$ relation of $\Lambda$CDM haloes. For each posterior sample, we compute the mean concentration $c_{200}^{\rm mean}(M_{200},z_{\rm l})$ at the lens redshift using the \citet{DiemerJoyce2019} $\mathrm{c}$--$M$ relation as implemented in \textsc{Colossus} \citep{Diemer2018Colossus}. We then assign an additional log-normal weight
\begin{equation}
    w_{\rm cM} \propto \exp\left[
        -\frac{1}{2}
        \left(
            \frac{\log_{10} c_{200} - \log_{10} c_{200}^{\rm mean}(M_{200},z_{\rm l})}
                 {\sigma_{\log c}}
        \right)^{2}
    \right],
\end{equation}
with a scatter of $\sigma_{\log c} = 0.18$~dex, consistent with the intrinsic scatter found in $N$-body simulations (e.g. \citealt{Bullock2001,DiemerJoyce2019}). This procedure down-weights posterior samples with concentrations that are highly inconsistent with the expected $\mathrm{c}$–$M$ relation.

Finally, we compute two reweighted posteriors: a “lensing + $\mathrm{c}$–$M$ prior” posterior using $w_{\rm cM}$ alone, and a “lensing + kinematic + $\mathrm{c}$–$M$ prior” posterior using the combined weight $w_{\rm kin}\,w_{\rm cM}$. These reweighted posteriors, shown in Figure~\ref{fig:posterior} (in green and black, respectively), tighten the constraints on the halo mass and concentration while leaving the lensing observables fully consistent with the original “lensing-only” solution.

\begin{table*}
  \centering
  \caption{
    Posterior constraints for the different lens models considered in this work. 
    The first block lists the stellar+gNFW models under four inference configurations (lensing-only, lensing+kinematics, lensing with a concentration--mass prior, and the combination of both). 
    The second block shows results for the stellar+EPL model where we use EPL mass profile to represent dark matter component. 
    The third block reports the constraints for the pure-EPL model taken from \citet{Collett2014}. 
    Reported values are the median of the marginal posterior, and the quoted uncertainties correspond to the 68\% credible interval.
    We assume a Planck18 cosmology and a lens redshift $z=0.222$.
    }
    \label{tab:posterior}
  { \renewcommand{\arraystretch}{1.3}
  \begin{tabular}{lrrrrrrrr}
    \hline
    Model & $M_{\mathrm{200}}^{\mathrm{halo}} / 10^{12} M_\odot$ & $R_{\mathrm{s}}^{\mathrm{halo}}\,[\mathrm{arcsecond}]$ & $\gamma_{\mathrm{in}}^{\mathrm{halo}}$ & $q_{\mathrm{halo}}$ & $\nabla (M/L)$ & $M_\star / 10^{11} M_\odot$ & $\theta_{\mathrm{E}}^{\mathrm{SIS,s1}}$ & $\alpha_{\rm SPS}$ \\
    Lensing only & $7.26^{+3.36}_{-2.08}$ & $7.0^{+5.4}_{-2.2}$ & $0.89^{+0.22}_{-0.19}$ & $0.85^{+0.02}_{-0.03}$ & $-0.06^{+0.04}_{-0.06}$ & $4.23^{+0.36}_{-0.48}$ & $0.17^{+0.03}_{-0.03}$ & $0.91^{+0.17}_{-0.15}$ \\
    kinematics & $7.33^{+3.66}_{-1.77}$ & $6.7^{+5.1}_{-2.0}$ & $0.85^{+0.19}_{-0.16}$ & $0.85^{+0.02}_{-0.03}$ & $-0.06^{+0.04}_{-0.07}$ & $4.23^{+0.35}_{-0.48}$ & $0.16^{+0.02}_{-0.02}$ & $0.91^{+0.16}_{-0.15}$ \\
    c-M prior & $9.32^{+3.87}_{-2.96}$ & $12.5^{+4.6}_{-4.4}$ & $1.11^{+0.14}_{-0.15}$ & $0.84^{+0.02}_{-0.02}$ & $-0.04^{+0.03}_{-0.04}$ & $4.38^{+0.30}_{-0.40}$ & $0.18^{+0.04}_{-0.03}$ & $0.95^{+0.16}_{-0.14}$ \\
    kinematics + c-M & $11.09^{+3.74}_{-3.20}$ & $12.6^{+4.9}_{-4.7}$ & $1.04^{+0.10}_{-0.14}$ & $0.84^{+0.02}_{-0.02}$ & $-0.03^{+0.02}_{-0.04}$ & $4.40^{+0.25}_{-0.39}$ & $0.16^{+0.02}_{-0.02}$ & $0.95^{+0.16}_{-0.14}$ \\
    \hline
      & $\theta_{\mathrm{E}}^{\mathrm{halo}}$ & $\gamma_{\mathrm{halo}}$ & $q_{\mathrm{halo}}$ & $\gamma_{\mathrm{shear}}$ & $\nabla (M/L)$ & $M_\star / 10^{11} M_\odot$ & $\theta_{\mathrm{E}}^{\mathrm{SIS,s1}}$ &  \\
    MGE + EPL & $0.534^{+0.077}_{-0.083}$ & $1.606^{+0.065}_{-0.062}$ & $0.886^{+0.016}_{-0.017}$ & $-0.016^{+0.039}_{-0.029}$ & $-0.225^{+0.056}_{-0.060}$ & $2.74^{+0.30}_{-0.27}$ & $0.145^{+0.031}_{-0.032}$ & -- \\
    \hline
     \cite{Collett2014} & $\theta_{\mathrm{E}}$ & $\gamma$ & $q$ & $\gamma_{\mathrm{shear}}$ & $\nabla (M/L)$ & $M_\star / 10^{11} M_\odot$ & $\theta_{\mathrm{E}}^{\mathrm{SIS,s1}}$ & $\eta$\\
    EPL & $1.397^{+0.001}_{-0.001}$ & $2.027^{+0.023}_{-0.025}$ & $0.946^{+0.009}_{-0.005}$ & $-0.069^{+0.002}_{-0.003}$ & -- & -- & $0.161^{+0.025}_{-0.021}$ & $1.405^{+0.014}_{-0.016}$ \\
    \hline
  \end{tabular}
  }
\end{table*}

\begin{figure*}
    \centering
    \includegraphics[width=0.8\textwidth]{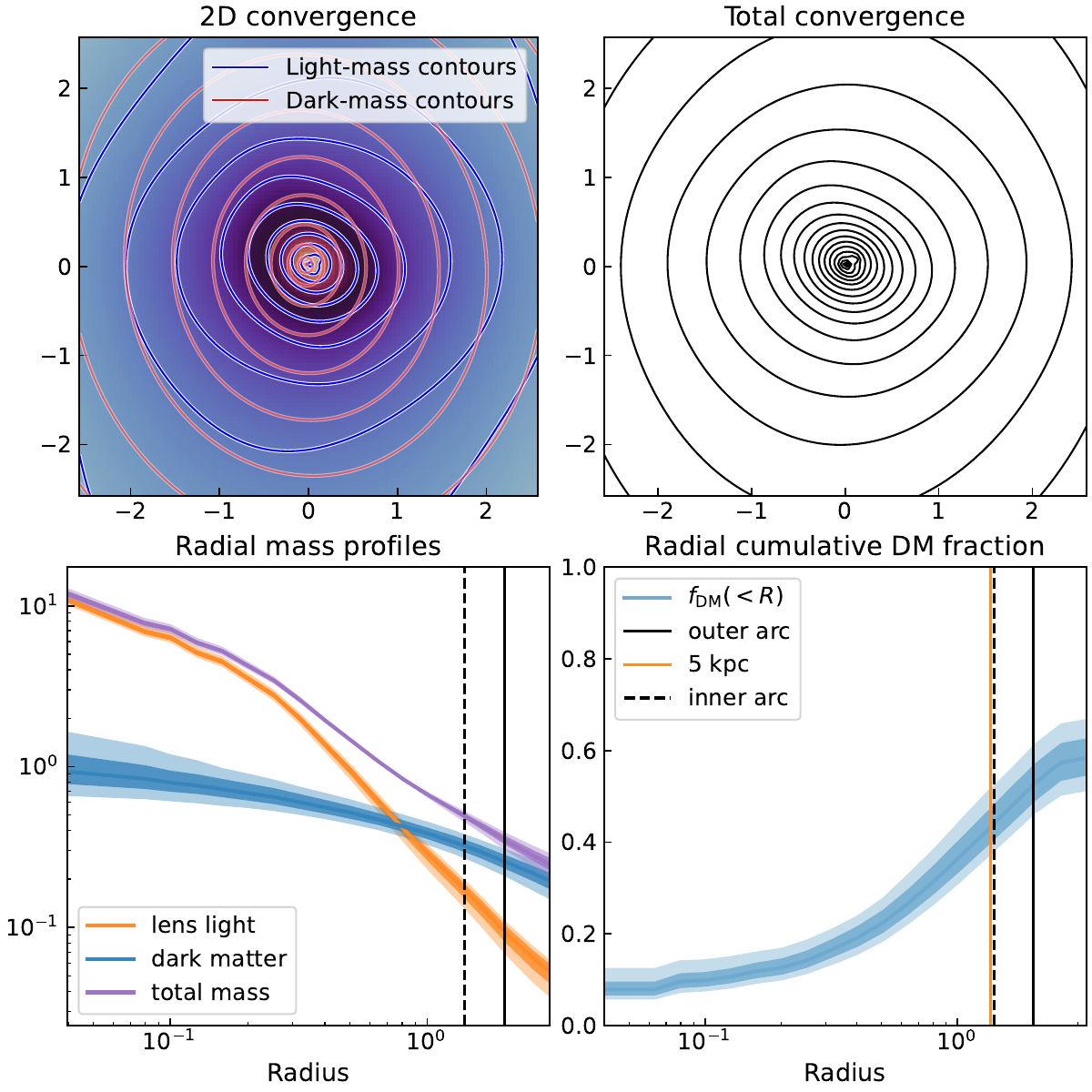}
    \caption{Mass decomposition of the lens model. 
    \textbf{Top-left:} Two-dimensional convergence map of the median lens model, where the total mass is shown in colour scale and the stellar (blue) and dark matter (red) components are shown as overlaid contours. 
    \textbf{Top-right:} Contours of the total convergence alone. 
    \textbf{Bottom-left:} Radial mass profiles of the stellar, dark matter, and total components. The solid curves show the median profiles, while the shaded bands indicate the 68\% and 95\% credible regions derived from 500 posterior samples. 
    \textbf{Bottom-right:} Cumulative dark matter fraction $f_{\rm DM}(<R)$ as a function of radius, with median (solid) and 68\%/95\% credible regions (shaded). The vertical dashed line marks the inner arc, while the solid black and orange lines indicate the outer arc and $R=5\,\mathrm{kpc}$, respectively.
    }
    \label{fig:dmfraction}
\end{figure*}

\begin{figure}
    \centering
    \includegraphics[width=0.4\textwidth]{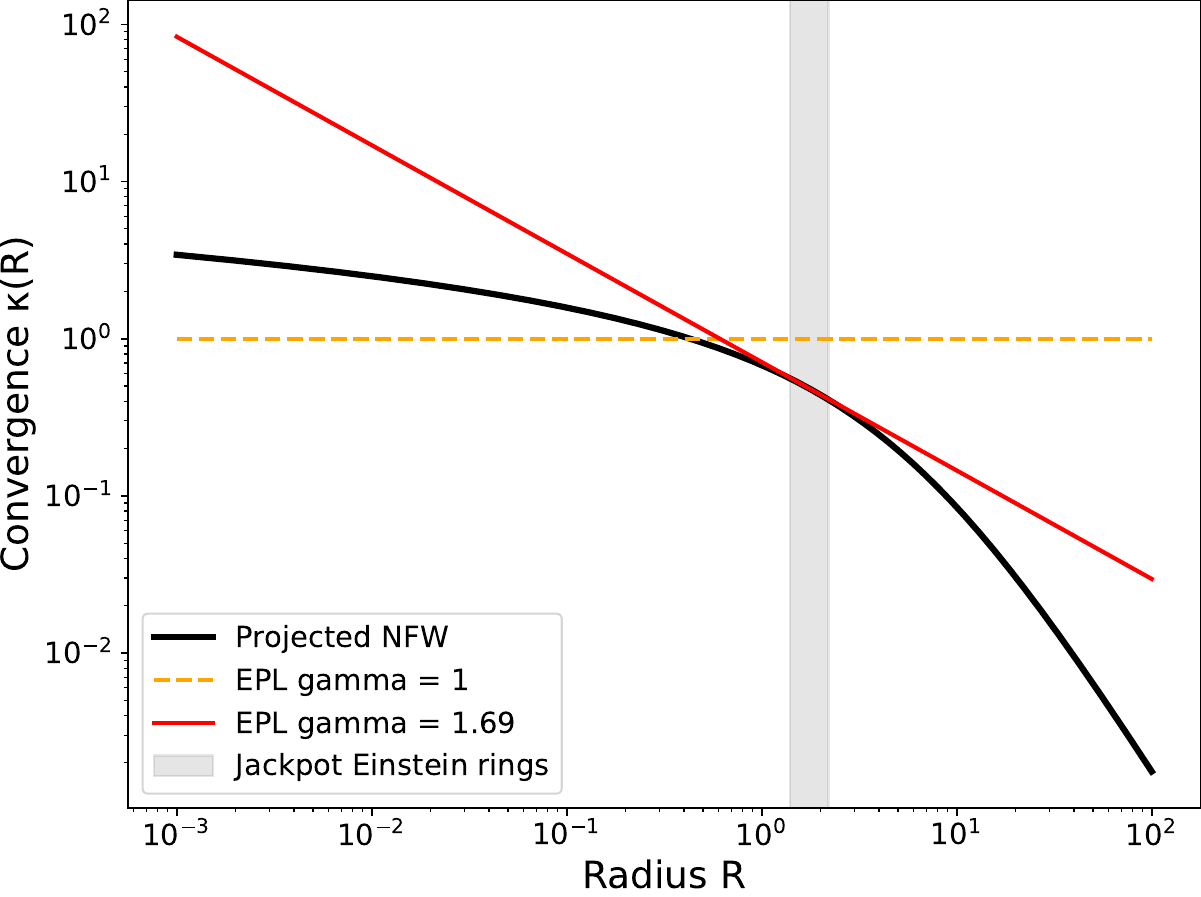}
    \caption{
    Projected surface--mass density profiles of an NFW halo and three power--law (EPL) models.  
    The black curve shows the projected NFW which has an Einstein radius of 0.53 arcseconds. All EPL models are normalised to have the same Einstein radius, $R_{\rm E}=0.53^{\prime\prime}$. The orange dashed curve shows an EPL model with $\gamma = 1.00$. The red curve shows the EPL profile with slope $\gamma_{\rm ring}$, obtained by matching the logarithmic slope of the NFW profile over the radial range $1.4$--$2.2^{\prime\prime}$ (where the two Einstein rings located at). 
    }
    \label{fig:NFW_EPL}
\end{figure}

\section{Lens model result}
\label{sec:lensmodl}
Figure \ref{fig:lensmodel} shows the lens model result, displaying the expectation values of each quantity taken over all the samples (i.e. the model image is evaluated at each sample and the mean is taken over these rather than evaluating the model at the mean of each parameter of the model). This avoids the projection effect whereby the marginal posterior median may lie outside the typical set (e.g. for strongly curved “banana-shaped’’ posteriors). We successfully modelled the lens image to the noise level, and focused both sources. 

Figure~\ref{fig:posterior} shows the posterior distribution of the stellar component plus gNFW dark matter model. Table~\ref{tab:posterior} summarises the marginalised posterior constraints for our fiducial stellar+gNFW model, the stellar+EPL model, and the pure-EPL model.

\subsection{Stellar component}
For our lensing-only model, the median stellar mass is
$M_\star = 4.23^{+0.36}_{-0.48}\times10^{11}\,M_\odot$
(corresponding to $\log_{10} M_\star = 11.63^{+0.036}_{-0.052}$).
The radial stellar mass-to-light gradient is restricted by the prior to be non-positive, and the posterior peaks at zero, indicating that the data prefer an approximately flat $M/L$ profile, consistent with the assumptions adopted by \citet{Sonnenfeld2012} and \citet{Turner2024}.
Quantitatively, we obtain the one-sided 95\% credible bound
$\nabla(M/L) > -0.18$
for the lensing-only inference, tightening to
$\nabla(M/L) > -0.12$
when including both the kinematic reweighting and the $c$--$M$ prior.

We quantify the IMF normalisation using the IMF mismatch parameter
\begin{equation}
\alpha_{\rm SPS} \equiv \frac{M_\star}{M_{\star,{\rm Salpeter}}},
\end{equation}
where $M_{\star,{\rm Salpeter}}$ is the Salpeter-IMF stellar mass \citep{Salpeter1955} inferred from stellar population synthesis (SPS) as reported by \citet{Sonnenfeld2012}.
For the lensing-only inference we obtain
$\alpha_{\rm SPS}=0.91^{+0.17}_{-0.15}$,
while for the reweighted posterior including kinematics and the $c$--$M$ prior we find
$\alpha_{\rm SPS}=0.96^{+0.16}_{-0.14}$.
In both cases $\alpha_{\rm SPS}$ is consistent with the Salpeter IMF stellar mass of \citet{Sonnenfeld2012}$.$
For comparison, adopting a Chabrier-IMF SPS stellar mass \citep{Chabrier2003} of $\log_{10}(M_{\star,{\rm Chab}}/M_\odot)=11.40\pm0.06$ would be in tension with our lensing-only stellar mass at the $\sim3\sigma$ level.

\subsection{Dark matter}

\begin{figure*}
    \centering
    \includegraphics[width=0.8\textwidth]{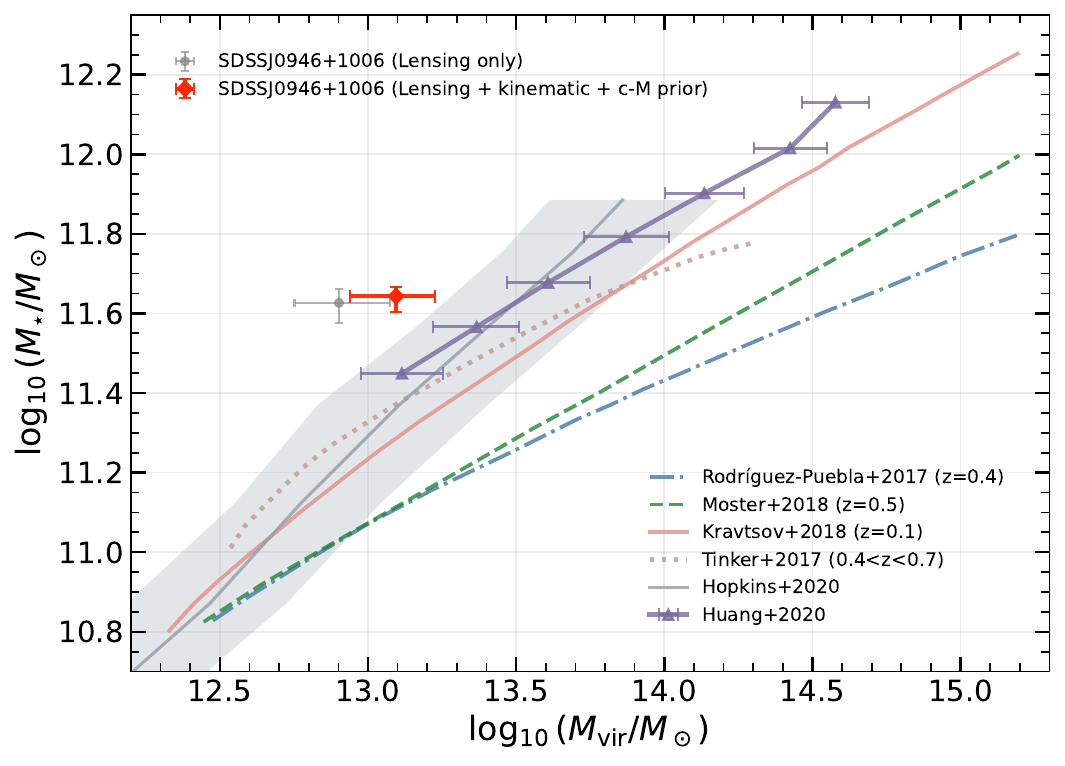}
    \caption{Stellar-halo mass relation in the $\log_{10}(M_{\rm vir})$--$\log_{10}(M_\star)$ plane.
    Our posterior constraints are shown for two inference setups: \emph{lensing only} and \emph{lensing + kinematics + $c$--$M$ prior}.
    For each setup, the marker denotes the posterior median, while the error bars indicate the central 68\% credible intervals.
    Halo masses are converted from $M_{200}$ to $M_{\rm vir}$ at $z=0.222$ using \textsc{Colossus} and the sampled $c_{200}$ values.
    Literature SHMRs are overplotted for comparison: \citet[$z=0.4$]{RodriguezPuebla2017},
    \citet[$z=0.5$]{Moster2018},
    \citet[$z=0.1$]{Kravtsov2018}, \citet[$0.4<z<0.7$]{Tinker2017}, \citet{Huang2018}, and \citet{Hopkins2020}; the shaded band indicates the uncertainty envelope reported in those works.
    }
    \label{fig:SHMR}
\end{figure*}

The dark matter profile is less well constrained than the stellar component in the lensing-only posterior, as expected: the dark halo is constrained only through the adopted mass-profile parametrisation and the lensing observables. As is typical, the inner slope of the gNFW profile is strongly degenerate with the halo scale radius (see Figure \ref{fig:full_posterior}). 
The marginalised posterior for the scale radius resembles an asymmetrically truncated Gaussian, with 
$R_{\mathrm{s}} = 7.0^{+5.4}_{-2.2}\,\mathrm{kpc}$. 
The median value is close to ten times the galaxy's half-light radius ($5^{\prime\prime}$), a value commonly adopted in previous studies \citep[e.g.][]{Melo2025}.

The lensing-only constraint on the inner slope is 
$\gamma_{\mathrm{in}} = 0.89^{+0.22}_{-0.19}$, 
which is broadly consistent with an NFW profile, but only weakly constrained. 
The constraining power of the lensing-only posterior is similar to that obtained by \citet{Sonnenfeld2012} and \citet{Turner2024} (at the $\pm 0.2$ level), but our result suggests that the halo inner slope is not as steep as reported in those studies. 
Unlike their analyses, however, we relax the assumptions on both the stellar $M/L$ gradient and the dark matter halo scale radius. 

\subsection{Comparison with the star+EPL model.}
When adopting a model more similar to \citet{Sonnenfeld2012}, in which the dark matter is described by an EPL profile, we obtain an inner slope of
$\gamma = 1.61^{+0.065}_{-0.062}$,
which is fully consistent with their measurement, but with roughly three times lower uncertainty.
Therefore, our star+EPL model reproduces the earlier conclusion of a steep inner profile.
However, the gNFW results indicate that this apparent steepness is largely an artefact of the power-law assumption. 
The tendency for EPL models (and gNFW models with very large scale radii, as seen in \citealt{Turner2024}) to prefer a steep inner slope is partially driven by limitations in their prior volume and model flexibility. 
In particular, a power-law mass model cannot access the low--inner-slope region of parameter space that a gNFW profile would otherwise allow. 
Moreover, the inner structure of a true NFW halo differs substantially from that of any single power law once projected. Figure~\ref{fig:NFW_EPL} illustrates the difference between a NFW halo and an EPL profile after projection to two dimensions. 
When a three-dimensional EPL profile is projected, its effective two-dimensional slope becomes shallower by one, which produces an almost flat projected profile. This behaviour is qualitatively different from a real NFW halo, whose projected convergence includes an additional $R/r_s$ dependence, so it still preserves some slope instead of being completely flat.

The red curve in Figure~\ref{fig:NFW_EPL} shows the ``equivalent'' EPL surface-density profile obtained by fitting a single power law to the projected NFW profile over the radial range $1.4$--$2.2^{\prime\prime}$ (where the lensed arcs are located). The resulting power-law slope is $\gamma \approx 1.69$, very close to the value recovered in our MGE+EPL model.
This demonstrates that a steep EPL slope can arise simply as the best single-power-law approximation to a projected NFW-like profile over the narrow radial range probed by the arcs, rather than requiring an intrinsically steep dark matter cusp.

\subsection{Total mass profile}
Figure~\ref{fig:dmfraction} highlights several features of the reconstructed mass distribution. Recently, the importance of including angular information in the mass distribution has been emphasised in several works \citep[e.g.][]{Powell2022, O'Riordan2024, Nightingale2024, Ballard2024, Enzi2025}. The top-right panel shows contours of the total convergence, which exhibit pronounced angular structure and clear deviations from simple elliptical symmetry. Inspecting the decomposition in the top-left panel reveals that this angular complexity arises from two contributions: higher-order, multipole-like structure in the lens light distribution itself, and a dark matter component whose ellipticity and position angle differ from those of the stellar mass. Because the angular structure is captured by physically distinct stellar and dark matter components, this star-plus–dark matter decomposition provides a natural, physically motivated baseline for measuring dark matter substructure in this system.

The blue stellar contours in the top-left panel, together with the total-convergence contours in the top-right panel, also reveal a small additional mass component near the main mass centre. This component has a radius of $\sim 0.035^{\prime\prime} (130\mathrm{pc})$, and mass of $7.4\times10^{8}\mathrm{M}\odot$. This extra light (and mass) component is likely associated with a satellite galaxy along the line of sight.

The radial profiles in the bottom panels show that the stellar and dark matter contributions are individually well constrained, and that their sum is extremely close to a single power law over the radial range probed, illustrating the usual bulge–halo conspiracy. Within \(R=5\,\mathrm{kpc}\), we infer a dark matter fraction \(f_{\rm DM}(<5\,\mathrm{kpc}) = 0.421^{+0.041}_{-0.035}\). This implies a stellar mass fraction \(f_\star \simeq 0.58\), which is consistent, within the uncertainties, with the mean stellar fraction \(0.46 \pm 0.13\) reported for SLACS lenses in \citep{Auger2010}. Compared with the population study of \citet{Shajib2021}, the Jackpot lens appears to have a higher inner dark-matter fraction. Our result is more consistent with the lensing and dynamical analysis of \citet{Oldham2018}, and with expectations from hydrodynamical simulations \citep[e.g.][]{xu2017}.

\subsection{Stellar-to-halo mass relation}

Our lens model constrains the stellar-to-halo mass ratio (SHMR). At the lens redshift $z=0.222$, our two-component model yields total halo masses of
$M_{200} \simeq 7.26^{+3.36}_{-2.08}\times10^{12}\,M_\odot$ (lensing only) and
$M_{200} \simeq 1.11^{+0.37}_{-0.32}\times10^{13}\,M_\odot$ (lensing+kinematics+$c$--$M$),
corresponding to $\log_{10}\!\left(M_{\rm halo}/M_\star\right)\simeq 1.25^{+0.16}_{-0.15}$ and $1.41\pm0.14$, respectively.
For comparison with the literature SHMRs in Fig. \ref{fig:SHMR}, we convert our halo masses from $M_{200}$ to $M_{\rm vir}$ at $z=0.222$ using \textsc{Colossus} and the sampled $c_{200}$ values. We caution, however, that the inferred total halo mass, M200, is more model-dependent because M200 is obtained by extrapolating the fitted halo model well beyond the radial range directly constrained by the lensing data, it is sensitive to the treatment of the scale radius ($\mathrm{R}_\mathrm{s}$).

Overall, our measurements are consistent with observationally derived SHMRs within the intrinsic scatter.
Across commonly used relations \citep[e.g.][]{RodriguezPuebla2017, Moster2018, Kravtsov2018, Tinker2017, Hopkins2020}, galaxies at $\log M_\star \gtrsim 11.3$ typically occupy haloes of $M_{\rm halo}\sim 10^{13}\,M_\odot$, with $\log_{10}(M_{\rm halo}/M_\star)\sim 1.3$ and an intrinsic scatter of order $\sim0.2$ dex.
Our lens lies slightly on the high-$M_{\rm halo}$ side at fixed $M_\star$, especially for the lensing+kinematics+$c$--$M$ setup.
The offset is modest, at the level of $\sim0.1$ dex relative to the typical expectation, and is comparable to the reported intrinsic scatter.
For a single system such an offset is not remarkable, though it is possible that the strong-lens selection function favours higher central surface densities or haloes oriented parallel to the line of sight \citep{Sonnenfeld2023Selection, Mandelbaum2009Selection, Tang2025LoS}.

\subsection{Third source}
\label{sec: thirdsource}
To test whether our composite lens model remains valid when propagating to the third source at \(z_3 = 5.975\), we extend the mass model to a three-plane configuration by adding a simple SIE component on the deflector plane associated with the second source. The centre of this SIE is determined by ray-tracing four conjugate image positions through the posterior-median mass model and taking the geometric centre of the traced conjugate positions. We then explore the Einstein radius \(\theta_{\mathrm{E}}\) and ellipticity of this additional SIE by minimising the separation of the third-source image pair in the final source plane. We effectively “focused’’ the third source to a minimum separation of \(\simeq 0.04^{\prime\prime}\).

This exercise is intended purely as a consistency check: given the highly complex morphology of the first and second sources, their associated mass distributions, as well as that of the third Lyman-\(\alpha\) emitter, are expected to be far more complicated than a single SIE. The third source is also nearly aligned with the first source but lies at a much larger projected distance from the second source. We find that varying the Einstein radius of the second source SIE has only a minor impact on the reconstructed position of the third source, changing its source-plane distance by less than 10 per cent. A more detailed investigation of how a more complex mass model for source~1 can further focus source~3 will be presented in Ballard et al.\ (in preparation).

\section{Systematics and limitation}
\label{sec:systematics}
The main difference between this work and previous studies that model galaxy-scale lenses with separate stellar and dark matter components is that we use the double-source-plane nature of the system to partially break the MST, rather than relying primarily on stellar kinematics. This reduces the systematic uncertainties associated with kinematic modelling, such as assumptions about velocity anisotropy, dynamical equilibrium, and aperture or seeing effects \citep{Toribio2026}. As shown in Fig.~\ref{fig:posterior}, including the kinematic information provides only a modest improvement in the constraints on the main mass parameters. We therefore conclude that, for the Jackpot, the DSPL geometry provides constraining power that is comparable to that from the kinematics. For systems where the lens is at higher redshift and has a small apparent size, the DSPL information is often more practical than stellar kinematics.

\subsection{Substructure}

A localised residual feature in the Jackpot has been interpreted as evidence for a substructure perturber in \citep{Vegetti2010}, although it has also been argued to be associated with a luminous satellite/bright structure \citep{Qiuhan2025}.

We do not explicitly model dark matter substructure in this work. Our main constraints are driven by the large-scale deflection field required to reproduce the two Einstein rings, whereas substructure primarily affects small-scale image features. Since distinguishing between these scenarios requires a dedicated substructure analysis beyond the scope of this paper, we treat such perturbers as a secondary source of systematics rather than a dominant limitation of our global mass-profile inference.

\subsection{Mass on the first source plane}

Compared to the kinematic approach, the DSPL approach introduces its own systematics, most importantly the need to model the mass distribution on the first source plane. For the Jackpot lens, source~1 is nearly perfectly aligned with the main lens, so the inferred source~1 mass is strongly degenerate with the main-lens mass. In particular, the mass enclosed by the second Einstein ring is effectively the sum of the main-lens mass and the source~1 mass contribution. As a result, changes in the source~1 mass model can be absorbed by changes in the main-lens dark matter halo.

In our tests, when we model source~1 as an SIE, the posterior tends to prefer a more elliptical dark matter halo. This behaviour is consistent with a compensation between the assumed source~1 ellipticity and the main-lens halo shape, and likely reflects remaining freedom in the source~1 parametrisation. 

A related issue is that an MST-like freedom also exists for the source~1 mass distribution. This can, in principle, propagate into the inferred main-lens profile parameters, including the dark matter inner slope. In contrast, we expect the SHMR to be less sensitive, because $M_{200}$ is mainly constrained by the global halo model (and strengthened by the $c$--$M$ prior).

A potential way to further reduce the degeneracy associated with the source~1 mass is to measure its spatially resolved kinematics. This requires high angular resolution integral-field spectroscopy, for example the \textit{JWST}/NIRSpec IFU, or future ELT IFU instruments such as HARMONI and METIS. However, simple Jeans modelling may not be well suited for this task. In DSPL systems, source~1 is often at relatively high redshift and may have a complex assembly history. For the Jackpot, source~1 shows two bright components, which is suggestive of a merger. In such a case the system may not be in dynamical equilibrium. The morphology is also clearly non-axisymmetric, which adds further complexity and makes common assumptions in Jeans analyses less reliable.

\subsection{Projection effect of the dark matter halo}

The halo axis ratio measured in this work is the projected axis ratio on the sky, and it does not uniquely determine the intrinsic three-dimensional halo shape. Strong lensing constrains the two-dimensional mass distribution on the sky, and it does not determine the three-dimensional shape of the halo or its orientation with respect to the line of sight. Therefore, the axis ratio reported here should be interpreted as the projected halo axis ratio. For the Jackpot in particular, the lens light shows clear tidal features, suggesting ongoing or recent gravitational interaction with a galaxy on the top right of the left panel of fig.~\ref{fig:rgbimg}. In this case, the inferred position angle of the dark matter halo is broadly consistent with the direction of the putative interaction. However, we do not expect most DSPLs exhibit such obvious tidal features, so this interpretation may not be general.

A practical way forward is to use a large sample of DSPLs and constrain the intrinsic halo-shape distribution statistically. In a population analysis, one can describe the distribution of triaxial haloes with some assumptions on their intrinsic alignments, generate the predicted distribution of projected axis ratios, and fit it to the data using forward modelling. The result from large population of DSPLs can be used to further study the intrinsic alignments and its correlations with the large scale structure.

\section{Discussion and conclusion}
\label{sec:discussion}

In this work, we model the Jackpot DSPL with a composite mass distribution consisting of star plus dark matter. The lens light is described by multiple non-concentric elliptical Gaussian components. We map light to stellar mass using a free global mass-to-light ratio and allow for a radial $M/L$ gradient. The dark matter halo is modelled with an elliptical gNFW profile. To accelerate the calculation of deflection angles, we approximate the gNFW halo with an MGE, while keeping all gNFW parameters free in the inference.

The main difference between this work and previous studies that model galaxy-scale lenses with separate stellar and dark matter components is that we use the double-source-plane nature of the system to partially break the MST, rather than relying primarily on stellar kinematics. This reduces the systematic uncertainties associated with kinematic modelling, such as assumptions about velocity anisotropy, dynamical equilibrium, and aperture or seeing effects (but introduces new systematics as discussed in Section \ref{sec:systematics}). As shown in Fig.~\ref{fig:posterior}, including the kinematic information provides only a modest improvement in the constraints on the main mass parameters. We therefore conclude that, for the Jackpot, the DSPL geometry provides constraining power that is comparable to that from the kinematics. For systems where the lens is at higher redshift and has a small apparent size, the DSPL information is often more practical than stellar kinematics.

Our main results are:
\begin{itemize}
    \item The stellar component prefers a flat $M/L$ profile and is consistent with a Salpeter IMF. We infer a total stellar mass of
    $M_{\star} = 4.4^{+0.25}_{-0.39}\times 10^{11}\,M_{\odot}$.
    
    \item The dark matter halo is close to NFW, with inner slope
    $\gamma_{\rm in}^{\rm halo} = 1.04^{+0.10}_{-0.14}$.
    The projected axis ratio is
    $q_{\rm halo} = 0.84 \pm 0.02$.
    
    \item The halo mass is
    $M_{200}^{\rm halo} = 1.11^{+0.37}_{-0.32}\times 10^{13}\,M_{\odot}$,
    implying
    $\log_{10}\!\left(M_{200}/M_{\star}\right)=1.41^{+0.13}_{-0.14}$.

     \item This stellar-to-halo mass ratio is higher than the typical expectation from literature stellar--halo mass relation at similar $M_\star$ and redshift, by $\sim0.1$ dex, which is comparable to the reported intrinsic scatter.

     \item This work demonstrates that DSPLs provide constraints on the mass profile that are comparable in constraining power to those from stellar kinematics.
\end{itemize}

In conclusion, we fit a flexible star-plus-dark-matter mass model to the Jackpot lens. Despite allowing both a radial stellar $M/L$ gradient and a generalised NFW halo, the data prefers an approximately constant stellar mass-to-light ratio with a Salpeter-like IMF normalisation, and a dark matter halo consistent with NFW. These conclusions hold for the lensing-only inference and remain unchanged when incorporating external constraints. Together with results from \citet{Schuldt2019} and \citet{Melo2025}, our measurement adds to emerging evidence that the inner dark matter profiles of massive early-type strong lenses are typically close to NFW. 

% The fact that the Jackpot lies slightly on the higher side of the stellar-to-halo mass relation compared to typical literature results at similar redshift is likely due to selection effects. The system was identified in SDSS, whose targeting favours bright massive ETGs, and strong lensing selection more generally prefers galaxies with high central surface density. A larger, more homogeneous DSPL sample will be needed to determine whether this offset is primarily selection-driven or reflects an intrinsically high halo mass at fixed stellar mass for this system.

The Jackpot lies slightly on the higher side of literature stellar-to-halo mass relations, which may simply be a statistical fluctuation. However, since the system was identified in SDSS, whose targeting favours bright massive ETGs, and strong-lensing selection generally favours galaxies with high central surface density, some degree of selection bias may also contribute to this offset. A larger, more homogeneous DSPL sample will be needed to robustly measure the stellar-halo mass relation and account for selection-related effects

Such samples will become available soon. In the Euclid Quick Data Release~Q1, four galaxy-scale DSPL candidates have already been reported, and yield forecasts predict that the full Euclid survey could discover of order $10^{3}$ DSPLs \citep{Li2025b}. The Chinese Space Station Telescope \citep{Cao2024} and the Nancy Grace Roman Space Telescope \citep{Spergel2015Roman} will provide strong-lensing samples on a similar scale. Together, these surveys will enable population-level DSPL analyses that can statistically recover the intrinsic distribution of the dark matter halo properties. With a population level of dark matter halo properties and stellar initial mass function measurement, we will have more knowledge on the mass profile of the galaxy-scale strong lenses. This will be very useful for solving the mass-sheet degeneracy and for overcoming selection effects in strong lensing cosmology \citep{Li2024, Li2025a, Erickson2025}

\section*{Acknowledgements}
We would like to thank Alessandro Sonnenfeld for the many useful discussions that enriched this work.

Numerical computations were done on the Sciama High Performance Compute (HPC) cluster, which is supported by the ICG, SEPNet, and the University of Portsmouth.

This work has received funding from the European Research Council (ERC) under the European Union's Horizon 2020 research and innovation program (LensEra: grant agreement No 945536). TC is funded by the Royal Society through a University Research Fellowship. DJB acknowledges the support of the Australian Research 
Council Discovery Project DP230101775.

GG thanks Pietro Bergamini and Amata Mercurio for the discussions on \texttt{pPXF}.

For the purpose of open access, the authors have applied a Creative Commons Attribution (CC BY) license to any Author Accepted Manuscript version arising.

%%%%%%%%%%%%%%%%%%%%%%%%%%%%%%%%%%%%%%%%%%%%%%%%%%
\section{Data Availability}
The HST data used in this work are publicly available from the Mikulski Archive for Space Telescopes (MAST; \url{https://mast.stsci.edu/}).
The \texttt{Herculens} code is publicly available at \url{https://github.com/Herculens/herculens.git}.
The scripts used to generate the data products and figures are available at \url{https://github.com/astroskylee/Jackpot_GNFW.git}.
The posterior samples and MCMC chains generated in this study are available from the corresponding author upon reasonable request.



\bibliographystyle{mnras}
\bibliography{example} % if your bibtex file is called example.bib

%\begin{thebibliography}{99}
%\end{thebibliography}

%%%%%%%%%%%%%%%%%%%%%%%%%%%%%%%%%%%%%%%%%%%%%%%%%%

%%%%%%%%%%%%%%%%% APPENDICES %%%%%%%%%%%%%%%%%%%%%

\appendix
\twocolumn
\section{Jax lensing profile}
\label{app:jaxlensingprofile}
In this section we summarise the main equation of the deflection angles created by Gaussian components, in the next we will show how this can help with a large class of profiles.
\subsection{Deflection angle}

\cite{Shajib2019MGE} provided a efficient way to calculate the deflection angle for a single elliptical Gaussian convergence and a fast way to decompose any 2D mass profile into a sum of multiple of these Gaussian components.

Here we briefly summarise the main equation. The Elliptical Gaussian convergence is:
\begin{equation}
    \kappa(R) = \kappa_0 \exp\!\left(-\frac{R^2}{2\sigma^2}\right)
\end{equation}
where $R=\sqrt{q^2 x^2 + y^2}$, and $\kappa_0$ is the sum of the flux of this gaussian.

The deflection angle of a single mass profile is:
\begin{equation}
\begin{aligned}
\alpha^*(z)= & \kappa_0 \sigma \sqrt{\frac{2 \pi}{1-q^2}} \varsigma(z ; q),
\end{aligned}
\end{equation}
where $\alpha^*(z)$ is the conjugate of the complex deflection angle $\alpha(z) \equiv \alpha_x+\mathrm{i} \alpha_y$ \citep{Bourassa1973}.
\begin{equation}
\varsigma(z ; q)=\varpi\left(\frac{q z}{\sigma \sqrt{2\left(1-q^2\right)}} ; 1\right)-\varpi\left(\frac{q z}{\sigma \sqrt{2\left(1-q^2\right)}} ; q\right)
\end{equation}
where $\varpi(z ; q)=\exp \left(-z^2\right) \operatorname{erfi}(\mathrm{qx}+\mathrm{iy} / \mathrm{q})$. We can express $\varpi(z ; q)$ using the Faddeeva function $w_{\mathrm{F}}(z)$ as
\begin{equation}
\begin{gathered}
\varpi(z ; q)=\mathrm{e}^{-x^2-2 \mathrm{i} x y} \mathrm{e}^{y^2}-\mathrm{i} \exp \left[-x^2\left(1-q^2\right)-y^2\left(1 / q^2-1\right)\right] \\
\times w_{\mathrm{F}}(q x+\mathrm{i} y / q) .
\end{gathered}
\end{equation}
In {\tt Jax-Lensing-Profiles}, we have implemented our Faddeeva function using regionwise strategy of \citet{Zaghloul2017}.

\subsection{MGE decomposition}
\begin{figure}
    \centering
    \includegraphics[width=0.45\textwidth]{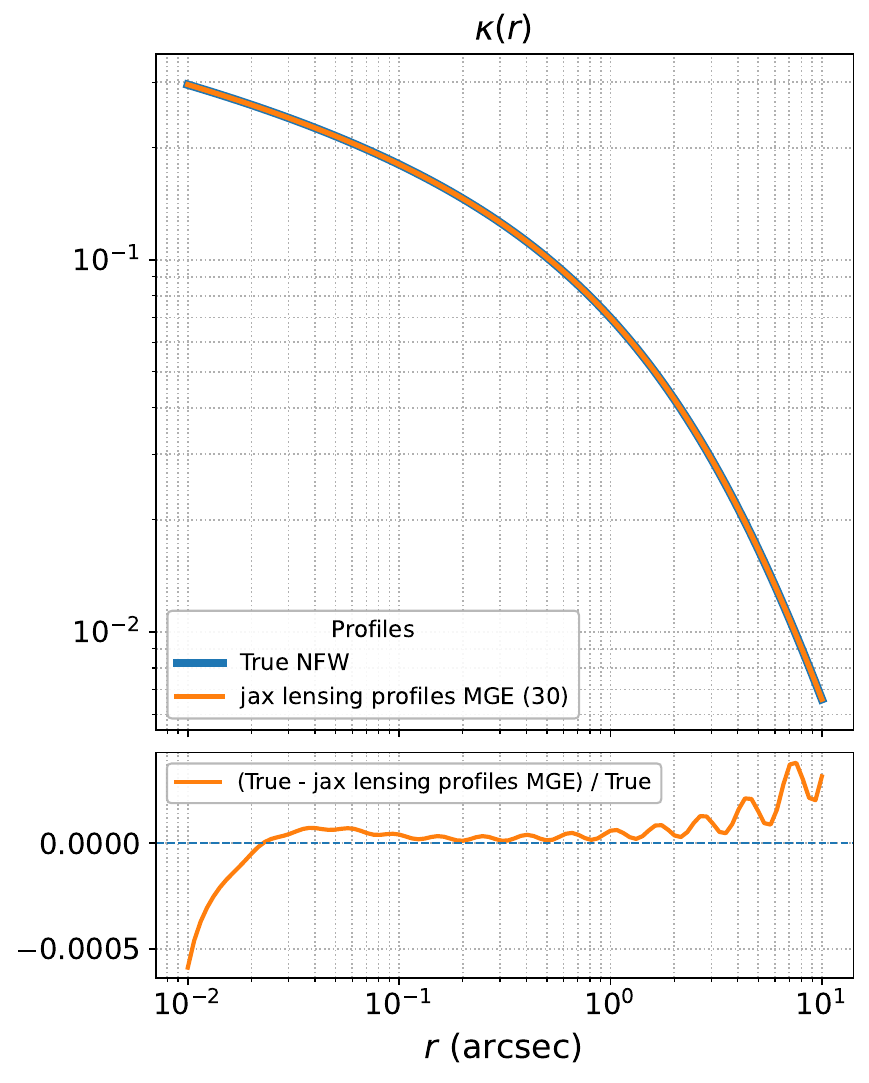}
    \caption{Illustration of approximating an NFW halo with a 30-component MGE. Top panel: radial convergence profile $\kappa(r)$ for the true NFW model (blue) and its MGE approximation (orange). Bottom panel: fractional difference relative to the true NFW profile.}

    \label{fig:3DMGE}
\end{figure}

\label{sec: MGE}
This method aims to approximate a function F(x) (in our case, the GNFW profile) as a sum of Gaussian components as:

\begin{equation}
\label{eq:MGE_sum}
F(x) \approx \sum_{n=0}^N A_n \exp \left(-\frac{x^2}{2 \sigma_n^2}\right)
\end{equation}

This equation is the discrete form of the following integral transform:
\begin{equation}
F(x) \equiv \int_0^{\infty} \frac{f(\sigma)}{\sqrt{2 \pi} \sigma} \exp \left(-\frac{x^2}{2 \sigma^2}\right) \mathrm{d} \sigma\
\label{eqa:14}
\end{equation}

In \cite{Shajib2019MGE}, equation \ref{eqa:14} can be Laplace-transformed by suitable change of variables. Then, the inverse transform of F(x) can be approximated by the Euler algorithm:

\begin{equation}
f(\sigma) \approx \sum_{n=0}^{2 P} \eta_n \operatorname{Re}\left[F\left(\sigma \chi_n\right)\right]
\end{equation}
where $A_n$ and $\sigma_n$ are respectively the amplitude and the standard deviation of the $n$-th Gaussian component. The weight and nodes are given by:
\begin{equation}
\begin{aligned}
& \chi_n=\left[\frac{2 P \log (10)}{3}+2 \pi \mathrm{i} n\right]^{1 / 2}, \\
& \eta_n=(-1)^n 2 \sqrt{2 \pi} 10^{P / 3} \xi_p, \\
& \xi_0=\frac{1}{2}, \quad \xi_n=1,1 \leq n \leq P, \quad \xi_{2 P}=\frac{1}{2^P}, \\
& \xi_{2 P-n}=\xi_{2 P-n+1}+2^{-P}\binom{P}{n}, 0<n<P .
\end{aligned}
\end{equation}
The P value here is related with the precision of the approximation. In our implementation, the code can automatically detect the jax environment. The P would be 13 for 32-bit floating point number and 28 for 64-bit floating point number. In this work, the P is set as 28 since our source reconstruction requires 64-bit jax environment.

So, by having a mass profile F(x) and choosing logarithmically spaced $\sigma_i$ for each gaussian component \citep[similar to ][]{Cappellari2002_MGE, Shajib2021}, one can compute the f($\sigma$) at each $\sigma_i$. The $A_n$ in equation 13 is then:
\begin{equation}
    A_n= w_n f\left(\sigma_n\right) \Delta(\log \sigma)_n / \sqrt{2 \pi}
\end{equation}

 The weights $w_n$ here is determined using the trapezoidal method with weights $w_1=0.5, w_n=1$ for $1<n<N, w_N=0.5$.

\subsection{Project gNFW to 2D}
Section~\ref{sec: MGE} presents an algorithm that decomposes a two-dimensional mass distribution into a sum of 2D Gaussians. This approach is ideal for profiles such as NFW, for which the 2D projection is analytic. However, the gNFW profile adopted here does not have a closed-form 2D projection, and evaluating it numerically is computationally expensive \citep[e.g.][]{Keeton2001}. We therefore start from the three-dimensional density profile, perform the MGE decomposition in 3D, and then project each Gaussian component analytically to obtain the corresponding 2D surface-mass density.

Concretely, for a spherical 3D Gaussian component
\begin{equation}
\rho_i(r)=\rho_{0,i}\exp\!\left(-\frac{r^2}{2\sigma_i^2}\right),
\end{equation}
its line-of-sight projection is
\begin{equation}
\Sigma_i(R)=\int_{-\infty}^{+\infty}\rho_i\!\left(\sqrt{R^2+z^2}\right)\,dz
= \rho_{0,i}\,\sqrt{2\pi}\,\sigma_i\,
\exp\!\left(-\frac{R^2}{2\sigma_i^2}\right),
\end{equation}
i.e.\ a 2D Gaussian with the same width $\sigma_i$ and a amplitude multiplied by the factor $\sqrt{2\pi}\sigma_i$. This makes the projection step trivial once the 3D MGE coefficients are known, and we can then compute deflection angles using the standard MGE machinery.

Figure~\ref{fig:3DMGE} illustrates that our \textit{jax lensing profiles} (MGE) approximation closely reproduces the true \textit{NFW} model. We also benchmarked our \texttt{JAX} implementation of \textit{MGE} against the \texttt{lenstronomy} implementation and found excellent agreement, with differences of order $\sim10^{-6}$ in the deflection angle and $\sim8\times10^{-3}$ in the convergence.

% The residual differences are small over the radial range of interest and lead to negligible changes in the predicted deflection field.

\section{Mass-sheet Transform in a DSPL system}

\begin{figure}
    \centering
    \includegraphics[width=1\linewidth]{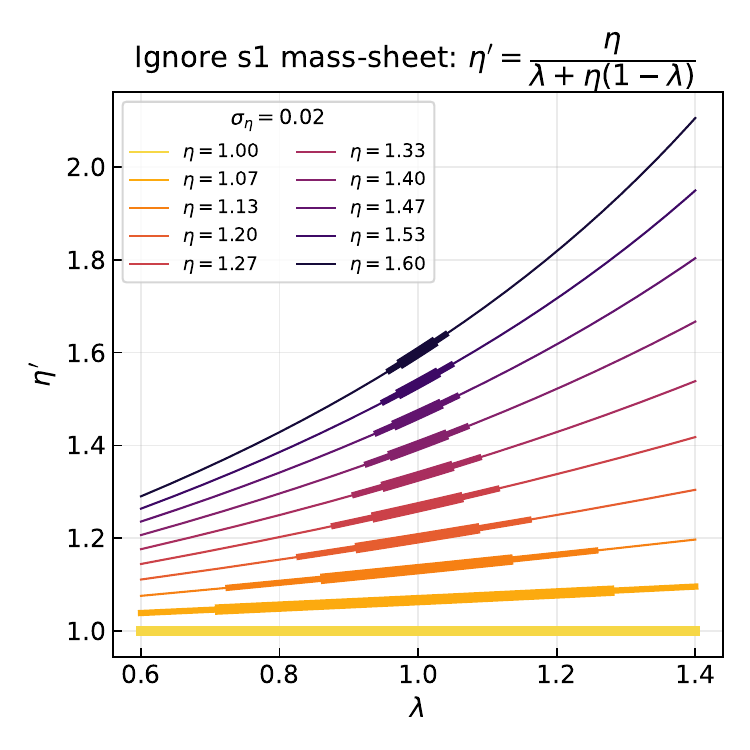}
    \caption{Relation between the mass-sheet parameter $\lambda$ and the resulting $\eta'$ for different input $\eta$ values, when we ignore the mass-sheet on the second lens plane (first source plane). We assume that the DSPL system has a true $\eta$, after MST becomes $\eta'$. We assume that a lens model on HST resolution can constrain $\eta'$ to 0.02 level, combining a known $\eta$ from cosmological model, we can have a constrain on $\lambda$. For each curve, the thick segment marks the $1\sigma$ region and the thinner segment marks the $2\sigma$ region.}
    \label{fig:mst_eta}
\end{figure}

\label{app:mst-dspl}

This appendix derives a general family of MSTs in the double-source-plane case. We apply a standard MST on plane~1, and on plane~2 we allow a uniform rescaling of the pre-existing deflection together with an isotropic mass sheet. At the same time, we allow the geometric distance ratio to change from $\eta$ to $\eta'$. The goal is to find parameters $(\lambda, v_3, K_2, \eta')$ such that the mapping to the far source plane (plane~3) is affected only by a global rescaling, preserving the lensing degeneracy at the level of image positions. This derivation is similar to \citet{Schneider2014}, but we adopt the alternative distance-ratio parametrisation $\eta \equiv 1/\beta$. In a general multi-plane system, our $\eta$ is defined with respect to the next plane in the mapping, while $\beta$ is conventionally defined with respect to the last plane; therefore $\eta=1/\beta$ holds only for the final $\eta$--$\beta$ pair (and is valid for the two-source DSPL case considered here). This choice matches our modelling convention: in our implementation we fit the inner Einstein ring first, and the deflection applied to the second source plane is evaluated by combining the deflection fields from the first and second lens planes with the geometry factor $\eta$.

We denote the angular coordinate on plane~1 (the main lens) by $\boldsymbol{\theta}_1$, and the mapped coordinates on plane~2 (which can host mass and the first source) and plane~3 (the farther source) by $\boldsymbol{\theta}_2$ and $\boldsymbol{\theta}_3$, respectively. The deflection fields on planes~1 and~2 are $\boldsymbol{\alpha}_1$ and $\boldsymbol{\alpha}_2$. The two-step mapping reads
\begin{align}
\boldsymbol{\theta}_2 &= \boldsymbol{\theta}_1 - \boldsymbol{\alpha}_1(\boldsymbol{\theta}_1),\\
\boldsymbol{\theta}_3 &= \boldsymbol{\theta}_1 - \eta\,\boldsymbol{\alpha}_1(\boldsymbol{\theta}_1)
                        - \boldsymbol{\alpha}_2(\boldsymbol{\theta}_2),
\end{align}
where $\eta$ is the geometry (distance--ratio) factor that weights the contribution of the plane~1 deflection as seen by plane~3.

We now apply a standard MST on plane~1,
\begin{align}
\boldsymbol{\alpha}_1'(\boldsymbol{\theta}_1)
  &= \lambda\,\boldsymbol{\alpha}_1(\boldsymbol{\theta}_1) + (1-\lambda)\,\boldsymbol{\theta}_1,
\qquad
\boldsymbol{\theta}_2'=\lambda\,\boldsymbol{\theta}_2,
\end{align}
so that the mapping to plane~2 is globally rescaled by $\lambda$.

On plane~2 we allow a uniform rescaling $v_3$ of the original deflection plus an isotropic sheet of strength $K_2$, and we simultaneously allow $\eta\rightarrow\eta'$,
\begin{align}
\boldsymbol{\alpha}_2'(\boldsymbol{\theta}_2')
  &= v_3\,\boldsymbol{\alpha}_2\!\left(\boldsymbol{\theta}_2'/\lambda\right)
   + K_2\,\boldsymbol{\theta}_2'
   = v_3\,\boldsymbol{\alpha}_2(\boldsymbol{\theta}_2)+K_2\,\boldsymbol{\theta}_2',
\\
\eta &\rightarrow \eta'.
\end{align}
The argument $\boldsymbol{\theta}_2'/\lambda=\boldsymbol{\theta}_2$ ensures that the rescaled deflection is evaluated at the corresponding pre-transform coordinate, while $\eta\to\eta'$ accommodates a change in the effective distance ratio (e.g.\ from cosmology or as a convenient reparametrisation).

We require that the mapping on the far source plane (plane~3) changes only by a global rescaling,
\begin{align}
\boldsymbol{\theta}_3' = v_3\,\boldsymbol{\theta}_3,
\end{align}
so that image positions remain degenerate up to this uniform factor. Using $\boldsymbol{\theta}_2'=\lambda(\boldsymbol{\theta}_1-\boldsymbol{\alpha}_1)$, the transformed mapping becomes
\begin{align}
\boldsymbol{\theta}_3'
= \bigl[1-\eta'(1-\lambda)-K_2\lambda\bigr]\boldsymbol{\theta}_1
  + \bigl[-\eta'\lambda+K_2\lambda\bigr]\boldsymbol{\alpha}_1(\boldsymbol{\theta}_1)
  - v_3\,\boldsymbol{\alpha}_2(\boldsymbol{\theta}_2),
\end{align}
which we match term-by-term to
\begin{align}
v_3\,\boldsymbol{\theta}_3
= v_3\Big[\boldsymbol{\theta}_1
          - \eta\,\boldsymbol{\alpha}_1(\boldsymbol{\theta}_1)
          - \boldsymbol{\alpha}_2(\boldsymbol{\theta}_2)\Big].
\end{align}
Equating coefficients of $\boldsymbol{\theta}_1$ and $\boldsymbol{\alpha}_1$ yields
\begin{align}
1-\eta'(1-\lambda)-K_2\lambda &= v_3, \label{eq:A}\\
-\eta'\lambda+K_2\lambda &= -v_3\,\eta. \label{eq:B}
\end{align}

For the generic case $\eta\neq 1$, Eqs.~\eqref{eq:A}--\eqref{eq:B} give
\begin{align}
\boxed{\,v_3=\frac{1-\eta'}{1-\eta}\,},\qquad
\boxed{\,K_2=\frac{\eta'\lambda-v_3\,\eta}{\lambda}
      \;=\;\eta'-\frac{\eta(1-\eta')}{\lambda(1-\eta)}\, }.
\end{align}
Thus, given $(\lambda,\eta,\eta')$, the degeneracy is fully characterized by $v_3$ (the global rescaling on plane~3) and the plane~2 sheet strength $K_2$.

These expressions recover the expected limits. For fixed cosmology ($\eta'=\eta$), we have $v_3=1$ and $K_2=\eta(\lambda-1)/\lambda$, corresponding to the usual ``MST on plane~1 plus an effective sheet on plane~2''. If instead we impose no sheet on plane~2 ($K_2=0$), then
\[
\eta'=\frac{\eta}{\lambda+\eta(1-\lambda)},\qquad
\lambda=\frac{\eta(1-\eta')}{\eta'(1-\eta)},\qquad
v_3=\frac{\lambda}{\lambda+\eta(1-\lambda)}.
\]
This provides the theoretical motivation for this work. Figure~\ref{fig:mst_eta} shows, for a fixed cosmology, how the constraint on the mass-sheet parameter $\lambda$ depends on the precision of the inferred distance-ratio parameter $\eta'$. For the Jackpot lens, where $\eta \simeq 1.4$, measuring $\eta'$ with an uncertainty of $\sigma_\eta \simeq 0.02$ corresponds to a constraint on $\lambda$ at the $\sim 5\%$ level.

It is sometimes useful to rewrite the relations in alternative forms. Expressing the change in geometry directly via the source rescaling $v_3$ gives
\begin{equation}
\boxed{\;\eta' = 1 - v_3\,(1-\eta)\;}, \qquad
\boxed{\;\eta = 1 - \frac{(1 - \eta')}{v_3}\;}.
\end{equation}
The convergence of the plane~2 can be written as
\begin{equation}
K_2 \;=\; 1 - v_3 \;+\; v_3\,\eta\,\frac{\lambda-1}{\lambda},
\end{equation}
and equivalently one may solve for $v_3$,
\begin{equation}
v_3 \;=\; \frac{1-K_2}{\,1-\eta+\eta/\lambda\,}
       \;=\; \frac{\lambda(1-K_2)}{\,\eta+\lambda(1-\eta)\,}.
\label{eq:v3_in_terms_of_K2}
\end{equation}
At fixed $(\lambda,\eta)$, Eq.~\eqref{eq:v3_in_terms_of_K2} shows that $K_2$ and the source 2 rescaling $v_3$ are related by a fractional-linear map: $K_2\!\to\!0$ reproduces the no--sheet limit above.

\section{Full posterior}

\begin{figure*}
    \centering
    \includegraphics[width=\textwidth]{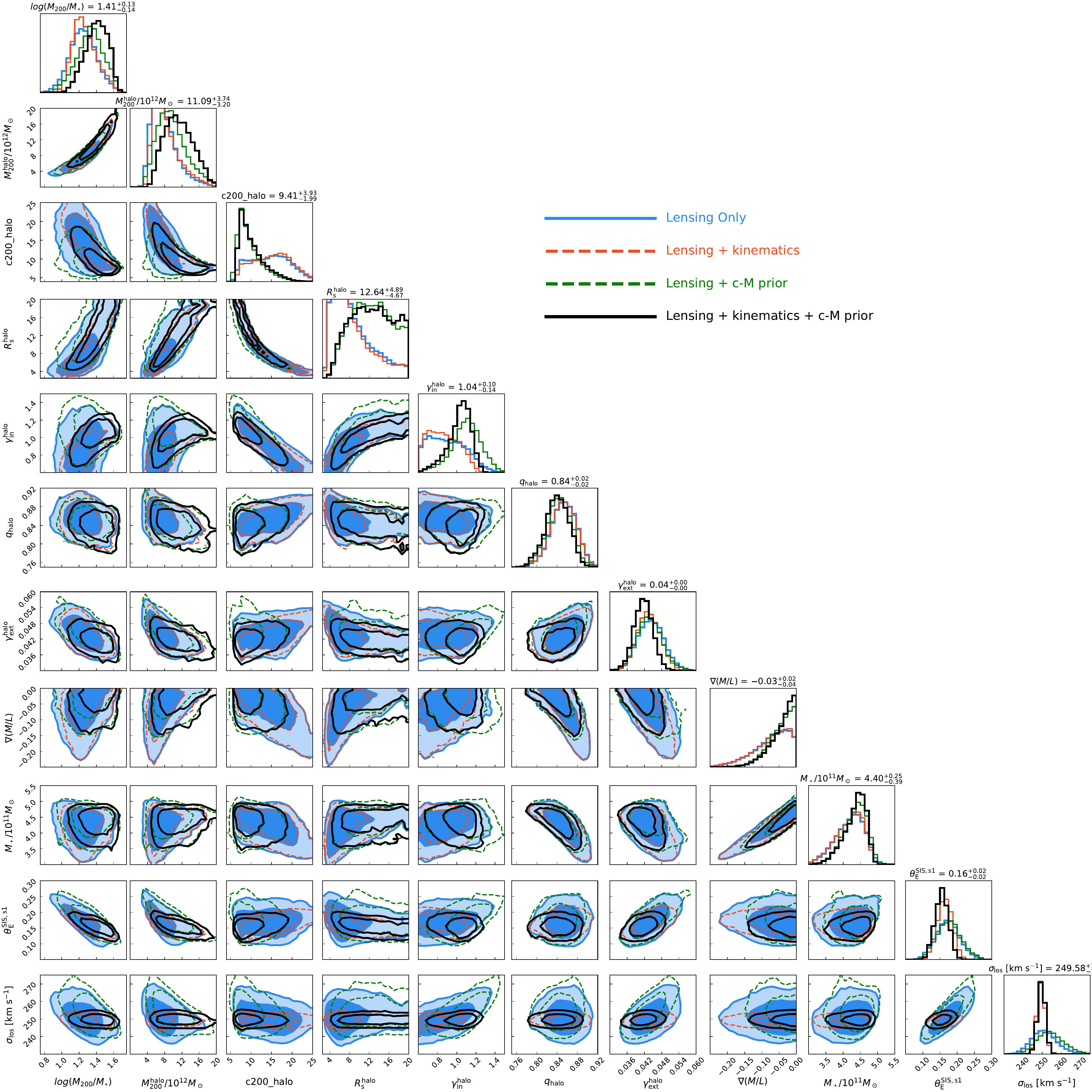}
    \caption{Posterior distributions for the composite star + gNFW halo model under different combinations of external information. Diagonal panels show the 1D marginalised posteriors; off--diagonal panels show the 2D posteriors with 68\% and 95\% credible contours. The parameters are $\log(M_{200}/M_\star)$ (stellar--to--halo mass ratio), $M_{\mathrm{200}}^{\mathrm{halo}} / 10^{12} M_\odot$ (halo mass), $c_{200}^{\mathrm{halo}}$ (halo concentration), $R_{\mathrm{s}}^{\mathrm{halo}}$ (halo scale radius), $\gamma_{\mathrm{in}}^{\mathrm{halo}}$ (inner logarithmic density slope), $q_{\mathrm{halo}}$ (projected axis ratio), $\gamma_{\mathrm{ext}}^{\mathrm{halo}}$ (external shear strength), $\nabla (M/L)$ (logarithmic radial gradient of the stellar mass--to--light ratio), $M_\star / 10^{11} M_\odot$ (total stellar mass), $\theta_{\mathrm{E}}^{\mathrm{SIS,s1}}$ (Einstein radius of the SIS component for source s1, in arcsec), and $\sigma_{\rm los}$ (aperture--averaged line--of--sight velocity dispersion, in km s$^{-1}$). Filled blue contours show the lensing--only inference; orange dashed lines show the posterior reweighted by the kinematic likelihood; green dashed lines include the concentration--mass prior; and solid black lines show the combined lensing + kinematic + concentration--mass constraints.
}
    \label{fig:full_posterior}
\end{figure*}

% Don't change these lines
\bsp	% typesetting comment
\label{lastpage}
\end{document}